\documentclass[aps,prl,twocolumn,superscriptaddress]{revtex4-2}

\usepackage{graphicx,epstopdf,siunitx,braket,xspace,amssymb,dsfont,bm,natbib,xfrac,nicefrac,miller,gensymb}
\RequirePackage[T1]{fontenc}
\RequirePackage[utf8]{inputenc}
\RequirePackage{lmodern}
\newcommand{\refeq}[1]{Eq.\,(\ref{#1})}
\newcommand{\reftab}[1]{Tab.\,\ref{#1}}
\newcommand{\refsec}[1]{Sec.\,\ref{#1}\,}
\DeclareRobustCommand{\reffig}[1]{Fig.\,\ref{#1}\,}

\newcommand{\qdU}[0]{\ensuremath{\Ket{\downarrow\Uparrow}}\xspace}
\newcommand{\quD}[0]{\ensuremath{\Ket{\uparrow\Downarrow}}\xspace}
\newcommand{\quU}[0]{\ensuremath{\Ket{\uparrow\Uparrow}}\xspace}
\newcommand{\qdD}[0]{\ensuremath{\Ket{\downarrow\Downarrow}}\xspace}
\newcommand{\qd}[0]{\ensuremath{\Ket{\downarrow}}\xspace}
\newcommand{\qu}[0]{\ensuremath{\Ket{\uparrow}}\xspace}
\newcommand{\qD}[0]{\ensuremath{\Ket{\Downarrow}}\xspace}
\newcommand{\qU}[0]{\ensuremath{\Ket{\Uparrow}}\xspace}

\begin{document}

\title{An electrically-driven single-atom `flip-flop' qubit}

\author{Rostyslav Savytskyy}
\thanks{These authors contributed equally to the present work}
\author{Tim Botzem}
\thanks{These authors contributed equally to the present work}
\affiliation{School of Electrical Engineering \& Telecommunications, UNSW Sydney, Sydney, NSW 2052, Australia}
\affiliation{Centre for Quantum Computation \& Communication Technology}

\author{Irene Fernandez de Fuentes}
\affiliation{School of Electrical Engineering \& Telecommunications, UNSW Sydney, Sydney, NSW 2052, Australia}
\affiliation{Centre for Quantum Computation \& Communication Technology}
\author{Benjamin Joecker}
\affiliation{School of Electrical Engineering \& Telecommunications, UNSW Sydney, Sydney, NSW 2052, Australia}
\affiliation{Centre for Quantum Computation \& Communication Technology}
\author{Jarryd J. Pla}
\affiliation{School of Electrical Engineering \& Telecommunications, UNSW Sydney, Sydney, NSW 2052, Australia}
\author{Fay E. Hudson}
\affiliation{School of Electrical Engineering \& Telecommunications, UNSW Sydney, Sydney, NSW 2052, Australia}
\affiliation{Centre for Quantum Computation \& Communication Technology}
\author{Kohei M. Itoh}
\affiliation{School of Fundamental Science and Technology, Keio University, Kohoku-ku, Yokohama, Japan}
\author{Alexander M. Jakob}
\affiliation{School of Physics, University of Melbourne, Melbourne, VIC 3010, Australia}
\affiliation{Centre for Quantum Computation \& Communication Technology}
\author{Alexander M. Jakob}
\affiliation{School of Physics, University of Melbourne, Melbourne, VIC 3010, Australia}
\affiliation{Centre for Quantum Computation \& Communication Technology}
\author{Brett C. Johnson}
\affiliation{School of Physics, University of Melbourne, Melbourne, VIC 3010, Australia}
\affiliation{Centre for Quantum Computation \& Communication Technology}
\author{David N. Jamieson}
\affiliation{School of Physics, University of Melbourne, Melbourne, VIC 3010, Australia}
\affiliation{Centre for Quantum Computation \& Communication Technology}
\author{Andrew S. Dzurak}
\affiliation{School of Electrical Engineering \& Telecommunications, UNSW Sydney, Sydney, NSW 2052, Australia}
\affiliation{Centre for Quantum Computation \& Communication Technology}
\author{Andrea Morello}
\email{To whom correspondence should be addressed; E-mail: a.morello@unsw.edu.au}
\affiliation{School of Electrical Engineering \& Telecommunications, UNSW Sydney, Sydney, NSW 2052, Australia}
\affiliation{Centre for Quantum Computation \& Communication Technology}

\date{}


\begin{abstract}
	The spins of atoms and atom-like systems are among the most coherent objects in which to store quantum information. However, the need to address them using oscillating magnetic fields hinders their integration with quantum electronic devices. Here we circumvent this hurdle by operating a single-atom `flip-flop' qubit in silicon, where quantum information is encoded in the electron-nuclear states of a phosphorus donor. The qubit is controlled using local electric fields at microwave frequencies, produced within a metal-oxide-semiconductor device. The electrical drive is mediated by the modulation of the electron-nuclear hyperfine coupling, a method that can be extended to many other atomic and molecular systems. These results pave the way to the construction of solid-state quantum processors where dense arrays of atoms can be controlled using only local electric fields.
\end{abstract}

\maketitle 

A century ago, understanding atoms' electronic structure and optical spectra was one of the first successes of the emerging theory of quantum mechanics. Today, atoms and atom-like systems constitute the backbone of coherent quantum technologies \cite{heinrich2021quantum}, providing well-defined states to encode quantum information, act as quantum sensors, or interface between light and matter. Their spin degree of freedom \cite{awschalom2013quantum} is most often used in quantum information processing because of its long coherence time, which can stretch to hours for atomic nuclei \cite{saeedi2013room,zhong2015optically}.

Moving from proof of principle demonstrations to functional quantum processors requires strategies to engineer multi-qubit interactions, and integrate atom-based spin qubits with control and interfacing electronics \cite{vandersypen2017interfacing}. There, the necessity to apply oscillating magnetic fields to control the atom's spin poses significant engineering problems. Magnetic fields oscillating at radiofrequency (RF, for nuclear spins) or microwaves (MW, for electron spins) cannot be localized or shielded at the nanoscale, and their delivery is accompanied by large power dissipation, difficult to reconcile with the cryogenic operation of most quantum devices. Therefore, many spin-based quantum processors based upon `artificial atoms' (quantum dots) rely instead on electric fields for qubit control, through a technique called electrically-driven spin resonance (EDSR) \cite{nowack2007coherent,pioro2008electrically}.

The longer coherence time of natural atoms and atom-like systems is accompanied by a reduced sensitivity to electric fields, making EDSR more challenging. EDSR was demonstrated in ensembles of color centers in silicon carbide \cite{Klimov2014} and molecular magnets \cite{liu2021quantum} for electron spins, and ensembles of donors in silicon for nuclear spins \cite{sigillito2017all}. At the single-atom level, coherent electrical control was demonstrated in scanning tunneling microscope experiments, \cite{Yang2019b}, single-atom molecular magnets \cite{Thiele2014} and high-spin donor nuclei \cite{asaad2020coherent}. Still lacking is a method to perform EDSR with single atoms, at MW frequencies, in a platform that enables easy integration with control electronics and large-scale manufacturing.

Here we report the coherent electrical control of a new type of spin qubit, formed by the anti-parallel states of the electron and nuclear spins of a single $^{31}$P atom in silicon, thus called `flip-flop' qubit \cite{tosi_silicon_2017}. The MW electric field produced by a local gate electrode induces coherent quantum transitions between the flip-flop states via the modulation of the electron-nuclear hyperfine interaction $A$, which depends on the precise shape of the electron charge distribution. Local control of $A$ with baseband electrical pulses was already suggested in the seminal Kane proposal \cite{kane_silicon-based_1998} as a way to select a specific qubit to be operated within a global RF magnetic field \cite{laucht_electrically_2015}. Here, instead, we control the qubit directly with local MW electric stimuli.

\begin{figure*}[htbp]
	\centering
	\includegraphics{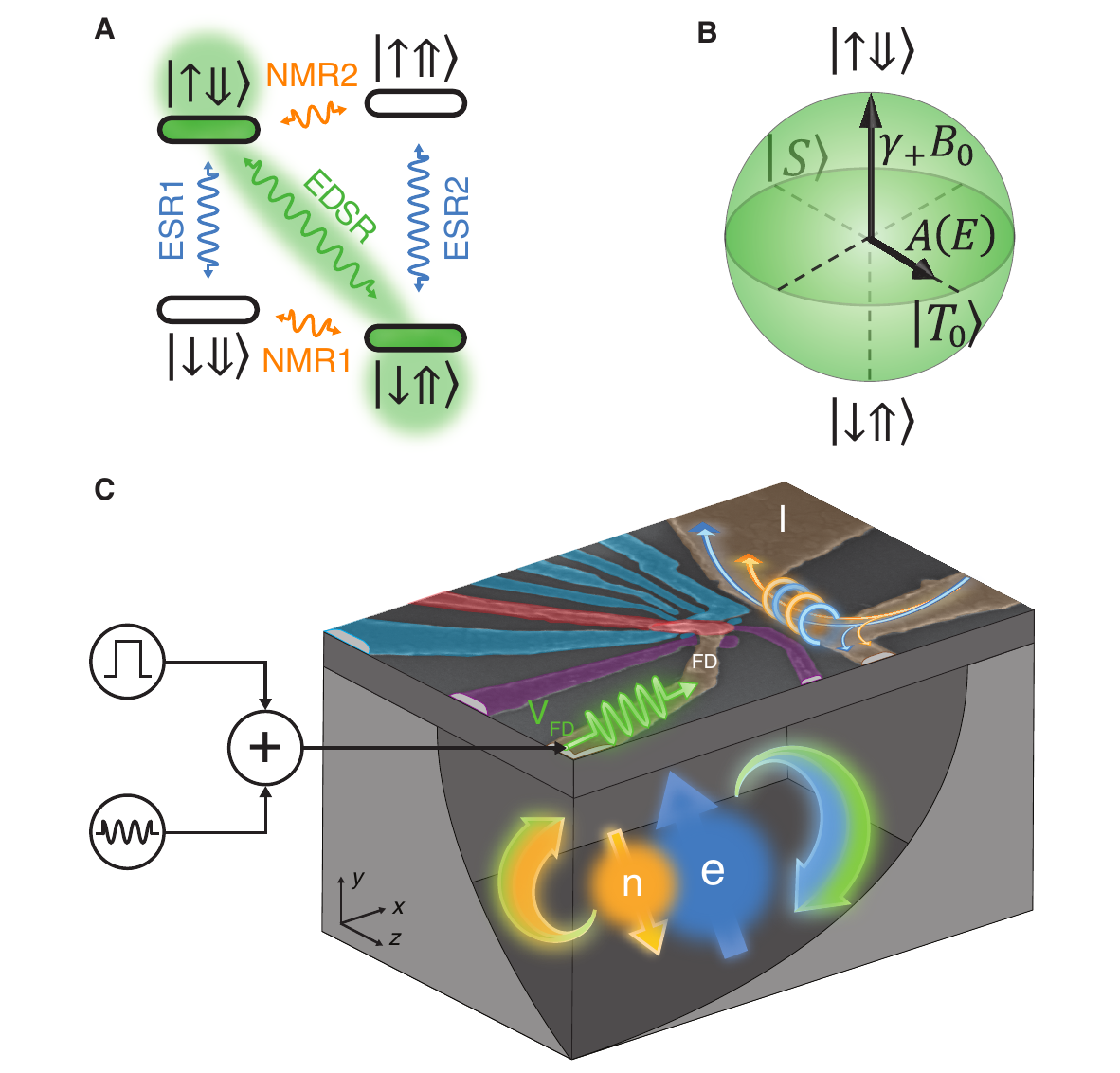}
	\caption{\textbf{Flip-flop qubit and device layout.} \textbf{(A)} Energy level diagram of $^{31}$P donor electron ($\ket{\uparrow},\ket{\downarrow}$) and nuclear ($\ket{\Uparrow},\ket{\Downarrow}$) spin states, in the presence of a static magnetic field $B_0 \sim 1$~T along the $z$ direction. ESR and NMR transitions are induced by oscillating magnetic fields. The flip-flop qubit is obtained by truncating the system to the $\ket{\downarrow\Uparrow},\ket{\uparrow\Downarrow}$ states, between which transitions are induced by electrically-driven spin resonance. \textbf{(B)} Bloch sphere representation of the flip-flop qubit. \textbf{(C)} False-color scanning electron microscopy image of the device, comprising a single electron transistor (cyan) to read out the electron spin, local gate electrodes (red and purple) to control the donor potential, and microwave antennas (brown) for electric (left, open-circuit) and magnetic (right, short-circuit) control of the donor spins. Here and elsewhere we use the color orange to represent properties related to the nuclear spin, blue for the electron spin, and green for the flip-flop qubit.}
	\label{fig:1}
\end{figure*}

The definition and operation of the flip-flop qubit can be understood on the basis of the spin Hamiltonian of the $^{31}$P donor system (see Supplementary Information S1 for details). In the neutral charge state, the spin degrees of freedom of the donor consist of a nucleus with spin $I=1/2$, gyromagnetic ratio $\gamma_\mathrm{n} = 17.23$~MHz/T and basis states $\ket{\Uparrow},\ket{\Downarrow}$, and a bound electron with spin $S=1/2$, gyromagnetic ratio $\gamma_\mathrm{e} = 27.97$~GHz/T and basis states $\ket{\uparrow},\ket{\downarrow}$. The electron and the nucleus are coupled by the Fermi contact hyperfine interaction $A\sim 100$~MHz. Placing the donor in a static magnetic field $B_0 \approx 1$~T results in an electron Zeeman splitting much larger than the nuclear Zeeman and the hyperfine energies. Therefore, the eigenstates of the system are approximately the tensor-product states \qdU, \qdD, \quD, \quU (\reffig{fig:1}A). 

The flip-flop qubit is defined as the two-dimensional subspace with basis states \qdU and \quD, shown in green in \reffig{fig:1}A. The qubit is thus described on a Bloch sphere where the \qdU, \quD states are the poles, and the singlet ($\ket{S}=(\ket{\downarrow\Uparrow}-\ket{\uparrow\Downarrow})/\sqrt{2}$) and unpolarized triplet ($\ket{T_0}=(\ket{\downarrow\Uparrow}+\ket{\uparrow\Downarrow})/\sqrt{2}$) states are at the equator (\reffig{fig:1}B), i.e.~they represent the eigenstates of the Pauli-$X$ operator in the flip-flop subspace. The energy term associated to the Pauli-$Z$ operator is $\gamma_+ B_0$, with $\gamma_+ = \gamma_{\rm e} + \gamma_{\rm n}$, while for the Pauli-$X$ it is simply the hyperfine coupling $A$, since $\ket{S}$ and $\ket{T_0}$ are the eigenstates of the hyperfine Hamiltonian in zero magnetic field.

From \reffig{fig:1}B we derive the flip-flop resonance frequency as $\epsilon_{\rm ff} = \sqrt{\left(\gamma_+B_0\right)^2 + A(E_{\rm dc})^2}$, where $E_{\rm dc}$ is the static electric field at the donor location. As in any qubit, transitions between the basis states are induced by a resonant modulation of the Pauli-$X$ term, which here is embodied by the hyperfine interaction. Therefore, the flip-flop Rabi frequency $2g_{\rm E}^{\rm ff} \equiv f_{\rm Rabi}^{\rm ff}$ depends upon the electric polarizability of the electron wavefunction, which is reflected in the Stark shift of the hyperfine coupling $\partial A(E)/\partial E$, yielding $f_{\rm Rabi}^{\rm ff} = (\partial A(E)/2\partial E)E_{\rm ac}$, where $E_{\rm ac}$ is an oscillating electric field and the factor 2 accounts for the rotating wave approximation. Unlike earlier examples of electrical drive in electron-nuclear systems \cite{Thiele2014,sigillito2017all}, the flip-flop transition does not require an anisotropic hyperfine interaction.

\begin{figure*}[htbp]
	\centering
	\includegraphics{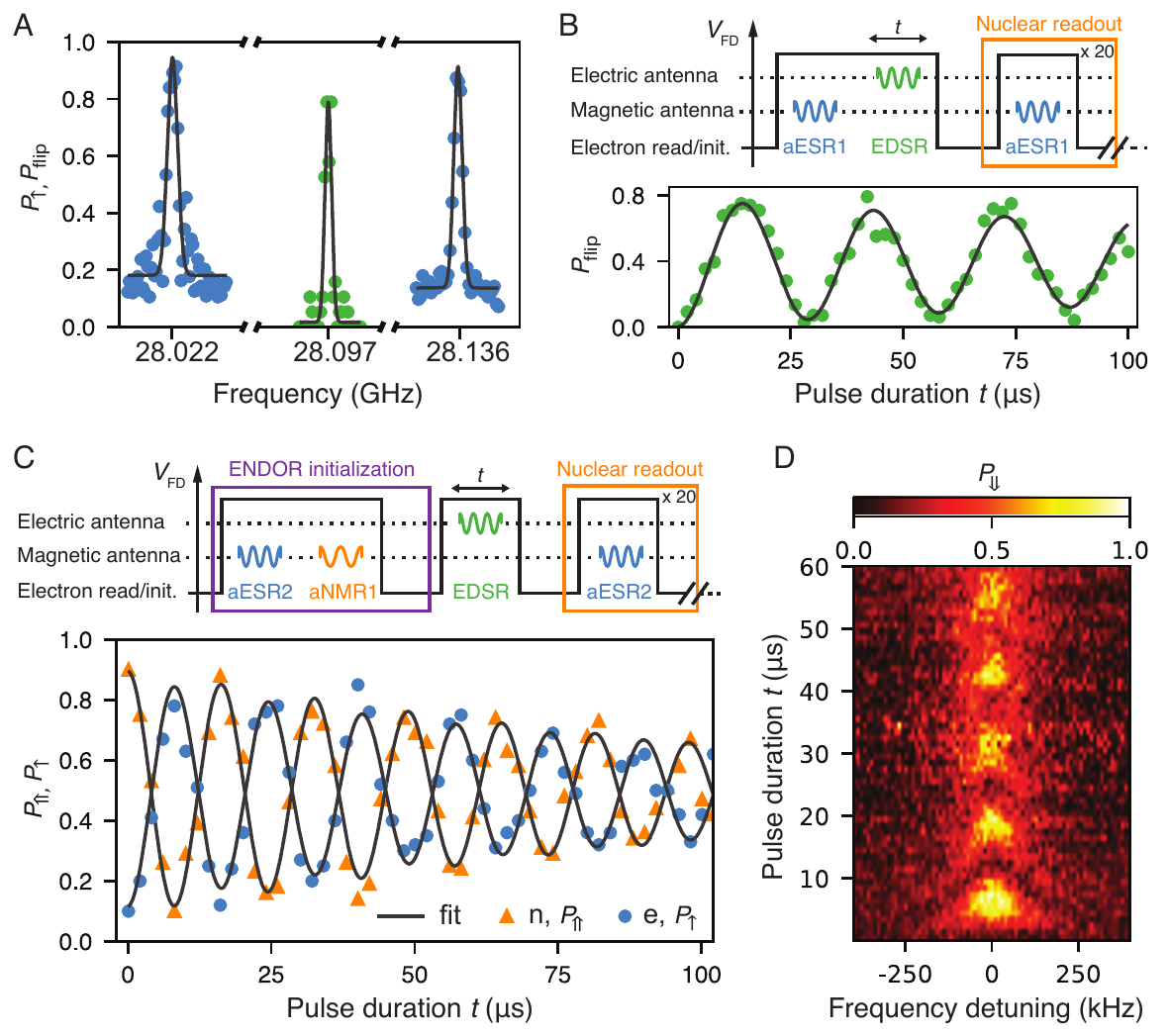}
	\caption{\textbf{Coherent electrical drive.} \textbf{(A)} The frequency spectrum shows two ESR peaks separated by $A = 114.1$~MHz. The flip-flop resonance peak is at $f_\mathrm{EDSR} = 28.0966$~GHz. \textbf{(B)} Coherent EDSR Rabi oscillation obtained by reading out the nuclear spin flip probability $P_\mathrm{flip}$. A schematic of the pulse sequence is shown on top. \textbf{(C)} Reading out the electron and nuclear spin-up proportions simultaneously highlights the anti-parallel flip-flop Rabi oscillations. A schematic of the pulse sequence is shown on top. \textbf{(D)} The Rabi chevron is mapped out by detuning the drive frequency around the resonance.}
	\label{fig:2}
\end{figure*}

To operate a flip-flop qubit, we fabricated a silicon metal-oxide-semiconductor (MOS) device as depicted in \reffig{fig:1}C. An ion-implanted $^{31}$P donor is placed under a set of electrostatic gates which control its electrochemical potential. An open-circuited electrical antenna, designed to drive EDSR, is connected to a high-frequency transmission line capable of carrying signals up to 40~GHz. A single-electron transistor is used for electron spin readout \cite{morello_single-shot_2010} and a short-circuited microwave antenna \cite{dehollain2012nanoscale} delivers oscillating magnetic fields at RF (for NMR) and MW (for ESR) frequencies (see Supplementary Information S2 for details).

We first acquire the ESR spectrum, which exhibits two resonances (one for each orientation of the nuclear spins) separated by $A = 114.1$~MHz. This value is close to $A = 117.53$~MHz observed in bulk experiments \cite{feher1959electron}. This suggest that the electron wavefunction of this donor, under the static electric fields used in this experiment, closely resembles that of a donor in the bulk. 

The flip-flop transition is found by applying a microwave tone to the electrical antenna, or fast donor (FD) gate (brown on the left in \reffig{fig:1}C) and measuring the resulting nuclear spin orientation \cite{pla_high-fidelity_2013}. Here the nuclear state is measured by applying an adiabatic frequency sweep \cite{laucht_high-fidelity_2014} of the MW drive on the magnetic antenna around the ESR1 resonance (adiabatic transitions are labeled with the prefix `a', e.g. aESR1 in this case), followed by electron spin readout. Detecting a $\ket{\uparrow}$ electron reveals that the nuclear spin is $\ket{\Downarrow}$. A high probability $P_{\rm flip}$ of the nuclear state changing from one shot to the next indicates the flip-flop resonance is being efficiently driven (see Supplementary Information for details). Note, however, that once the system is excited to the \quD state, the \qu electron is then replaced by a \qd one, bringing us outside the flip-flop qubit subspace. To prevent this, prior to each EDSR shot we apply an aESR1 pulse, which is off-resonant if the system is in the \qdU state but returns the system to \quD if it is in the \qdD state. We find a high nuclear spin flip probability at $f_\mathrm{EDSR} = 28.0966$~GHz (\reffig{fig:2}A), in agreement with the EDSR (flip-flop) transition frequency predicted from the values of $A, \gamma_{\rm e}, B_0$, measured independently. 

To demonstrate coherent electrical control of the flip-flop transition, we first perform an EDSR Rabi experiment by measuring the nuclear flip probability as a function of the duration of the electrical EDSR pulse (\reffig{fig:2}B), using the aESR1 pre-pulse to randomly initialize in the flip-flop subspace. Since the electron spin is itself read out as part of the nuclear readout process, we have the additional possibility of measuring the state of both spins and verifying the electron-nuclear flip-flop dynamics. We first employ an electron-nuclear double resonance (ENDOR) pulse sequence to deterministically initialize the system in the flip-flop ground state \qdU (see Materials and Methods for details). This ENDOR sequence comprises an aESR2 pulse, followed by an aNMR1 pulse and a subsequent electron readout, which initializes the electron to the $\ket{\downarrow}$ state (see \reffig{fig:1}A and \reffig{fig:2}C). If the system is in \qdU, the aESR2 pulse will excite it to \quU, the aNMR1 pulse will be off-resonant and the electron readout will initialize the system back to \qdU. If the system is in the \qdD prior to the ENDOR pulses, the aESR2 pulse is off-resonant and the aNMR1 pulse will flip the nucleus to \qdU.

Once initialized, we apply a resonant EDSR tone, which drives transitions from \qdU to \quD state, and first read out the electron spin \cite{morello_single-shot_2010}. Subsequently, we reload an electron onto the donor and perform the nuclear spin readout \cite{pla_high-fidelity_2013} as described earlier (see \reffig{fig:1}A and pulse sequence in \reffig{fig:2}C). By repeating this sequence more than 20 times, we determine the electron $P(\uparrow)$ and nuclear $P(\Uparrow)$ spin-up proportions for each EDSR pulse duration. \reffig{fig:2}C shows the anti-parallel coherent drive of both electron and nuclear spins. 

By detuning the drive frequency of the EDSR pulse around the resonance, we map out the Rabi chevron pattern of the nuclear spin (see \reffig{fig:2}D). Here, we again initialize the system in the flip-flop ground state \qdU using the ENDOR pulse sequence. 

\begin{figure*}[htbp]
	\centering
	\includegraphics{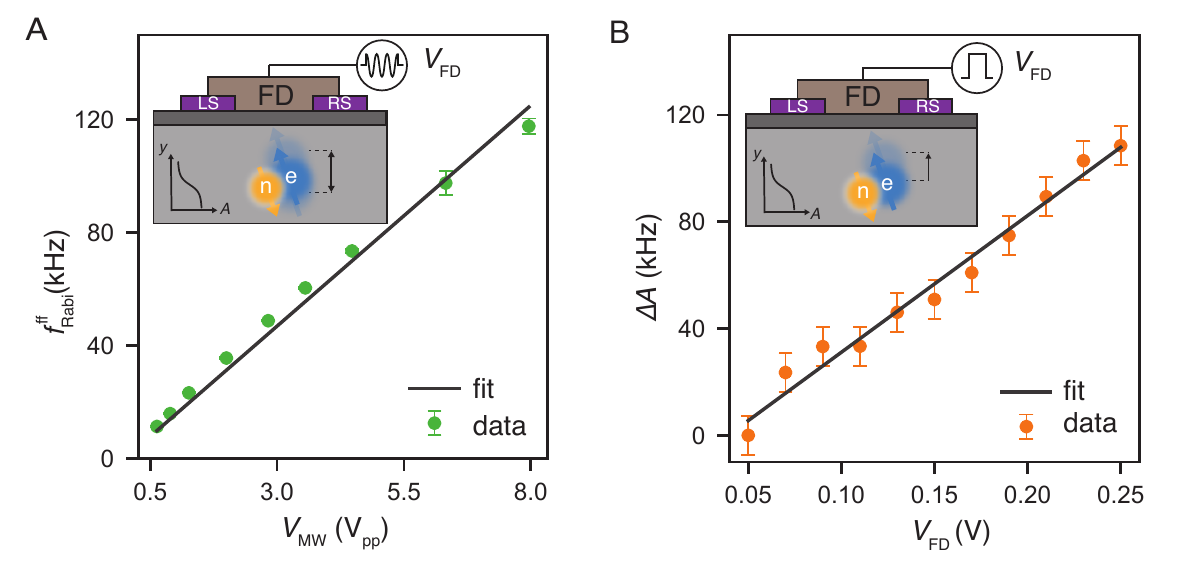}
	\caption{\textbf{Electrical drive via hyperfine modulation.} \textbf{(A)} Linear dependence of the flip-flop Rabi frequencies on on the amplitude at the voltage out put by the microwave source. \textbf{(B)} Stark shift of the hyperfine coupling produced by a DC voltage applied to the fast donor gate, as extracted from the shift of the NMR1 resonance frequency. An independent calibration of the line attenuation at MW confirms that the flip-flop qubit is driven by dynamic modulation of the hyperfine coupling. }
	\label{fig:3}
\end{figure*}

We find a linear dependence of the Rabi frequency on the drive power for the flip-flop transition indicating we are within the rotating wave approximation (see \reffig{fig:3}A). For the highest MW driving power (22 dBm at the source, corresponding to 8 V$_{\rm pp}$), we reach a flip-flop Rabi frequency $f_\mathrm{Rabi}^{\rm ff} = 118.5(25)$~kHz, which is a factor 5 higher than the fastest single-nucleus Rabi frequency reported in the literature \cite{pla_high-fidelity_2013}.

It is in general difficult to estimate the precise value of the oscillating voltage at the tip of the electrical antenna, due to the strongly frequency-depend losses of the transmission line between the MW source and the device. Here, however, we can correlate $f_\mathrm{Rabi}^{\rm ff}$ with the Stark shift of the hyperfine coupling, which can be measured independently. We apply a DC voltage shift $\Delta V_{\rm FD}$ to the fast donor gate (the same used for EDSR), and measure the hyperfine Stark shift $\Delta A(\Delta V_{\rm FD})$ through the shift of the NMR1 resonance, $f_\mathrm{NMR1} = A/2 + \gamma_n B_0$ (\reffig{fig:3}A). A linear fit to the data yields $\partial A/\partial V_{\rm FD} = 512(26)$~kHz/V. The positive slope of $A(V_{\rm FD})$ disagrees with the expectation that a positive gate voltage should pull the electron away from the donor nucleus, reducing $A$ \cite{tosi_silicon_2017}. We have performed a capacitance-based triangulation of the donor position (See Supplementary Information) and found it is located next to one of the side confining gates. Therefore, the more positive voltage on the tip of the FD gate may have the effect of bringing the electron closer to the nucleus, in a sideways direction. A similarly positive value of $\partial A/\partial V$ was also observed in earlier experiments \cite{laucht_electrically_2015}.

\begin{figure*}[htbp]
	\centering
	\includegraphics{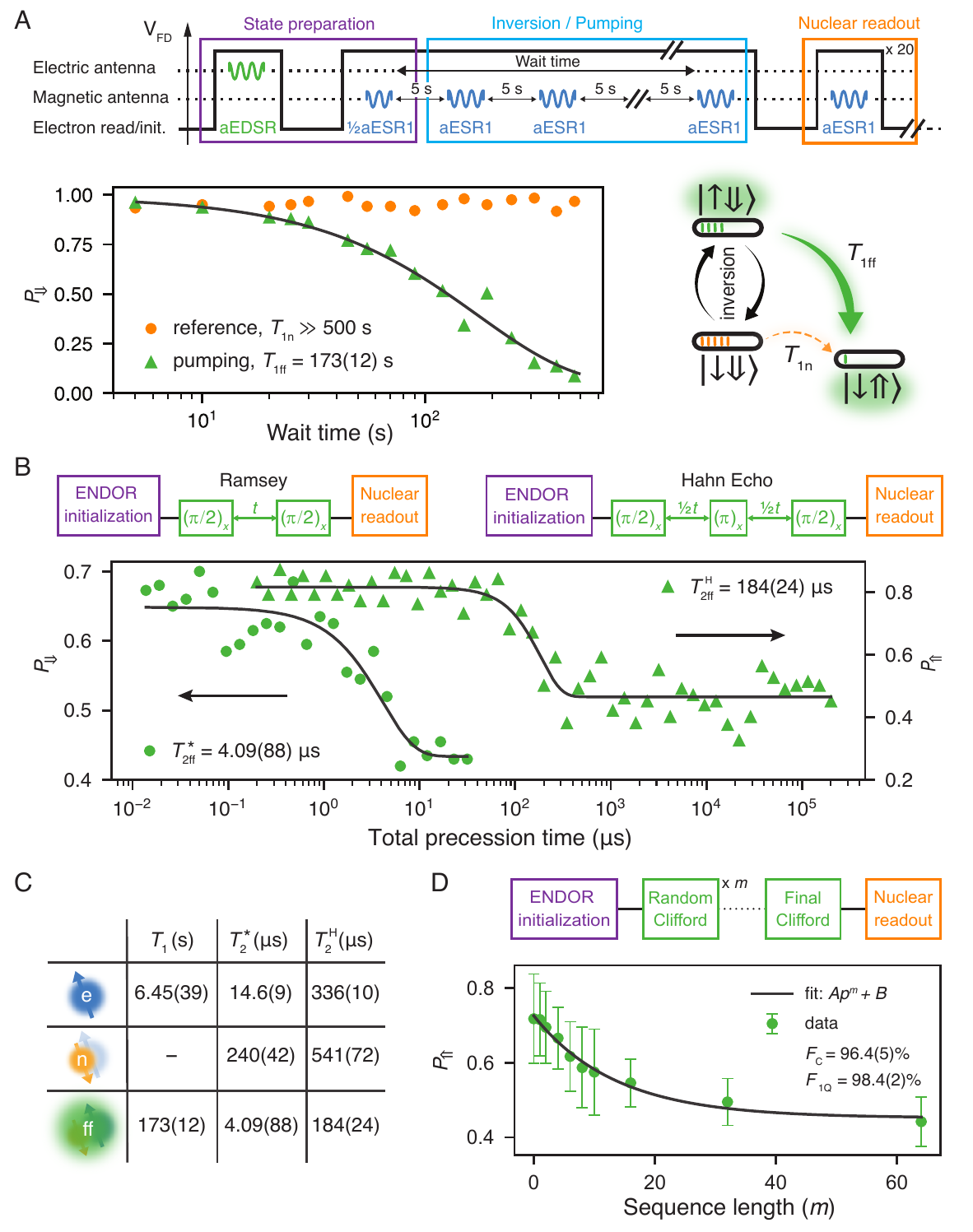}
	\caption{\textbf{Relaxation, coherence and gate fidelity.} \textbf{(A)} The flip-flop qubit relaxation time $T_{1\mathrm{ff}}$ = 173(12)~s is measured by initializing the donor in the $a\quD + b\qdD$ state with $|a|^2 \approx |b|^2 \approx 0.5$ and maintaining the \quD state population with adiabatic aESR1 inversion pulses applied every 5~s. \textbf{(B)} Fitting the Ramsey and the Hahn Echo decays using a exponential decays reveals $T_{\rm 2ff}^* = 4.09(88)$~$\mu$s (with exponent 1.28) and $T_{\rm 2ff}^{\rm H} = 184(24)$~$\mu$s (exponent 2), respectively. \textbf{(C)} Tabulated values of the relaxation and coherence times measured on the electron, nuclear, and flip-flop qubits. \textbf{(D)} Randomized benchmarking experiment for the flip-flop qubit, yielding an average one-qubit gate fidelity $\mathcal{F}_{\rm 1Q} = 98.4(2)\%$.}
	\label{fig:4}
\end{figure*}

Using the experimentally determined slopes $\partial f^\mathrm{ff}_\mathrm{Rabi}/\partial V_\mathrm{FD} $ and $\partial \Delta A/\partial V_\mathrm{FD}$ from \reffig{fig:3}A,B and the formula for the EDSR Rabi frequency, we determine that the total line attenuation between MW source and FD gate at 28~GHz is 18.1(5)~dB. This value is in good agreement with an independent estimate, 19.4(5)~dB, obtained by comparing the effect of a 100~Hz square pulse and the 28~GHz MW pulse on the broadening of the SET Coulomb peaks (see Supplementary Information). The slight discrepancy can be explained by the different capacitive couplings of the FD gate to the donor-bound electron and to the SET. Overall, this experiment confirms the validity of the assumption that the flip-flop qubit is driven by the electrical modulation of the hyperfine coupling.

Having demonstrated the coherent operation and readout of the flip-flop qubit, we proceed to measure its key performance metrics for quantum information processing, i.e. relaxation, coherence, and gate fidelities. 

From bulk experiments on $^{31}$P donors, the relaxation process within the flip-flop qubit subspace ($\ket{\uparrow\Downarrow}\rightarrow\ket{\downarrow\Uparrow}$) is known to be extremely slow, $T_{\rm 1ff} \approx 5$~hours \cite{feher1959electron}. Since the electron spin relaxation time ($\ket{\uparrow\Downarrow}\rightarrow\ket{\downarrow\Downarrow}$) is $T_{\rm 1e} = 6.45(39)$~s in this device, measuring $T_{\rm 1ff}$ requires saturating the ESR1 transition, while monitoring the escape of the system out of the $\ket{\downarrow}$ subspace via the flip-flop transition. We adopted the pumping scheme depicted in \reffig{fig:4}A: starting from the \qdD state, the donor is placed in a superposition state $a \qdD + b \quD$, with $|a|^2 \approx |b|^2 \approx 0.5$, using a slow frequency sweep (labeled $1/2$aESR1) calibrated to yield a 50\% probability of inverting the electron spin \cite{laucht_high-fidelity_2014}. After this, we apply aESR1 inversion pulses every 5 seconds to counteract the $T_{\rm 1e}$ process, and measure the nuclear state at the end. We obtain $T_{\rm 1ff} = 173(12)$~s, indeed much longer than $T_{\rm 1e}$. We also independently verified that, without repopulating the $\ket{\uparrow\Downarrow}$, the rate of nuclear spin flip $\ket{\downarrow\Downarrow}\rightarrow\ket{\downarrow\Uparrow}$ is immeasurably slow, $T_{\rm 1n}\gg 500$~s.

To investigate the coherence of the flip-flop qubit, we performed an on-resonance Ramsey experiment, where we initialize the system in the flip-flop ground state \qdU using the ENDOR sequence and apply two consecutive EDSR $\pi/2$-pulses separated by a varying time delay (see \reffig{fig:4}B). We obtained a pure dephasing time $T_{\rm 2ff}^* = 4.09(88)$~$\mu$s which is nearly a factor 4 lower than $T_{\rm 2e}^* = 14.6(9)$~$\mu$s for the donor-bound electron (\reffig{fig:4}C). 

A Hahn echo experiment is performed by applying a $\pi$-pulse halfway through the free evolution time, which decouples the qubit from slow noise and extends the coherence time to $T_{\rm 2ff}^{\rm H} = 184(24)$~$\mu$s (\reffig{fig:4}B). This value is approximately half the echo time for the electron, $T_{\rm 2e}^{\rm H} = 336(10)$~$\mu$s (\reffig{fig:4}C).

The microscopic origin of the flip-flop decoherence mechanisms, and their relation to the electron spin decoherence, is still under investigation. A key observation is that the EDSR pulses induce a transient shift of the resonance frequencies of up to 80~kHz, chiefly by affecting the electron gyromagnetic ratio (see Supplementary Information). The pulse-induced resonance shift (PIRS) depends on the power and duration of the pulse in a non-trivial way, and decays slowly after the pulse is turned off. Similar effects were reported earlier in the literature \cite{freer2017single,watson2018programmable,zwerver2021qubits} and attributed to heating or rectification effects, but remain poorly understood. 

Another key decoherence mechanism is the presence of residual $^{29}$Si nuclear spins in the substrate. Despite using an isotopically enriched $^{28}$Si material with 730~ppm residual $^{29}$Si concentration, we found the flip-flop and ESR resonances to be split in six well-resolved clusters of frequencies, indicating at least three $^{29}$Si nuclei coupled to the electron by $\sim 100$~kHz (see Supplementary Information). These nuclei flip as often as once per minute.

We measured the average one-qubit gate fidelities of the flip-flop qubit using the well-established methods of gate set tomography (GST) \cite{nielsen2021gate} and randomized benchmarking (RB) \cite{magesan2011scalable}. In both cases, the effect of $^{29}$Si nuclear spin flips is mitigated by sandwiching each gate sequence between spectrum scans to monitor the instantaneous resonance frequency. If the frequencies before and after the sequence are different, the measurement is discarded and repeated (see Supplementary text for details). Despite this precaution, the GST analysis reveals a strong deviation from a Markovian model, probably due to the residual effect of PIRS. Therefore, the GST one-qubit average fidelities $\mathcal{F}_{\rm 1Q} = 97.5\%-98.5\%$ are additionally verified by RB. GST also provides a value of $\mathcal{F}_{\rm SPAM} \approx 92\%$ for the state preparation and measurement (SPAM) fidelity of the \qdU state. A similar value, $\mathcal{F}_{\rm SPAM} = 90.9(6)$ was obtained through a direct measurement of the probability of preparing the \qdU with an ENDOR sequence (see Supplementary Information).

Randomized benchmarking determines the average gate fidelity by applying to the qubit, initialized in \qdU, a random sequence of Clifford gates with varying length $m$. In our compilation, the Clifford gates are composed of $\approx 2.233$ native gates $\in \{X, Y, \pm\sqrt{X}, \pm\sqrt{Y}\}$ on average. The last Clifford operation in each sequence is chosen such that the final state ideally returns to \qdU. The final state of the flip-flop qubit is measured by monitoring the probability $P_{\Uparrow}$ of finding the nuclear spin in the $\ket{\Uparrow}$ state. In the presence of gate errors, $P_{\Uparrow}$ decays as a function of sequence length $m$ and the average gate fidelity is determined from the decay rate (see \reffig{fig:4}C and Supplementary text for further details). We find an average Clifford gate fidelity $\mathcal{F}_\mathrm{C} = 96.4(5)\%$, which corresponds to an average native gate fidelity $\mathcal{F}_{\rm 1Q} = 98.4(2)\%$ (\reffig{fig:4}D).

Future experiments will focus on operating the flip-flop qubit in the regime of large electric dipole, where the wavefunction of the electron is equally shared between the donor and an interface quantum dot \cite{tosi_silicon_2017}. This regime is predicted to yield fast one-qubit (30~ns for a $\pi/2$ rotation) and two-qubit (40~ns for a $\sqrt{iSWAP}$) operations with fidelities well above 99\% under realistic noise conditions, with further improvements possible using optimal control schemes \cite{calderon2021fast}. The large induced electric dipole will allow the operation of donor qubit arrays with spacing of order 200~nm, with generous tolerances on the precise donor location, and dimensional compatibility with industry-standard manufacturing processes \cite{zwerver2021qubits,fischer2015}. The present setup did not allow reaching the large-dipole regime because of the presence of many other donors randomly implanted in the device: the large gate voltage swing necessary to move the electron away from the donor under study would unsettle the charge state of nearby donors (see Supplementary Information). The recent demonstration of deterministic single-ion implantation with 99.85\% confidence \cite{jakob2022deterministic} will eliminate this problem and unlock the full potential of the flip-flop qubit architecture.

In the present experiment, an on-chip antenna to deliver oscillating magnetic fields remains necessary in order to perform NMR control and ensure the system is prepared in the flip-flop subspace. Moving from $^{31}$P to heavier group-V donors such as $^{123}$Sb brings nuclei with $I>1/2$ whose electric quadrupole moment enables nuclear electric resonance (NER) \cite{asaad2020coherent}. Combined with the electrical control of the flip-flop transitions, NER will permit the full electrical control of the whole Hilbert space of all group-V donors other than $^{31}$P. The flip-flop drive can also be used to implement geometric two-qubit logic gates for nuclear spins, which have recently shown to yield universal quantum logic with fidelities above 99\% \cite{mkadzik2022precision}.

The results shown here already illustrate the broad applicability of the flip-flop qubit idea, even to atoms and atom-like systems that do not permit the creation of a large electric dipole, or do not possess anisotropic hyperfine couplings. For example, all-epitaxial donor devices fabricated with scanning probe lithography \cite{fricke2021coherent} do not allow the formation interface quantum dots, but their flip-flop states could be electrically controlled using the methods shown here. The flip-flop transition has already been used to hyperpolarize the nuclear spins on individual Cu atoms on a surface using a scanning tunneling microscope \cite{Yang2018}, and may be extended to coherently control the atoms' spins. Atoms \cite{Gilardoni2021} and atom-like defects \cite{Falk2014} in SiC possess significant electrical tunability of their electronic states, which may be exploited for flip-flop transitions in the presence of hyperfine-coupled nuclei. Molecular systems could permit even more tailored electrical responses, as already observed in recent ensemble experiments \cite{liu2021quantum}.

\section*{MATERIALS AND METHODS}

\label{sec:setup}

\textbf{Device fabrication}

\begin{figure*}[htbp]
\centering
\includegraphics{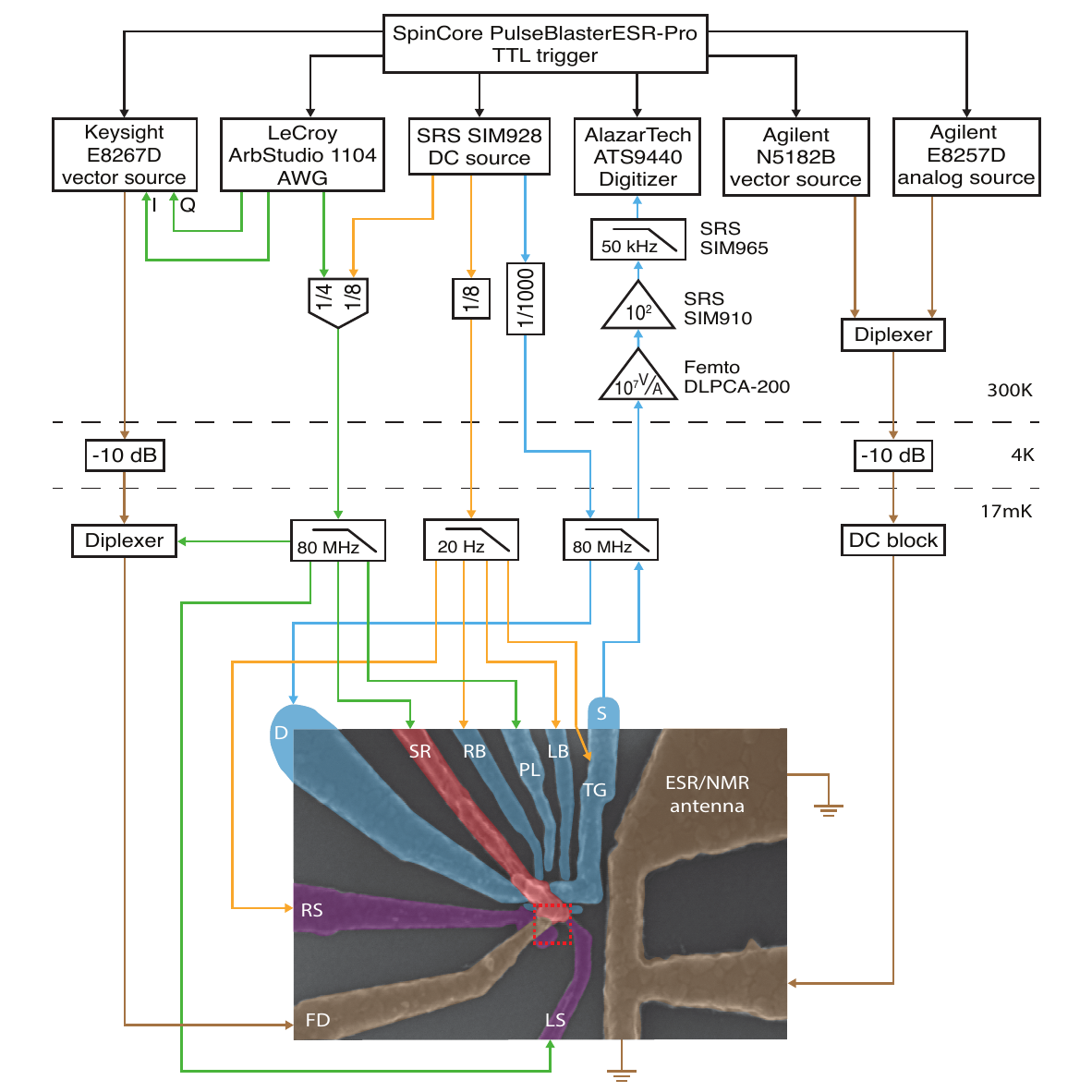}
\caption{\textbf{Experimental setup.} Wiring and instrumentation used to control and read out the donor spin qubit. The red dashed square defines the implantation region for this qubit device.}
\label{fig:setup}
\end{figure*}

The device under investigation is fabricated on an isotopically enriched $^{28}$Si wafer. We fabricate metal-oxide nanostructures in the proximity to the implantation area of the $^{31}$P donors (red dashed square in \reffig{fig:setup}) to manipulate and read out the donor spin qubit, similar to Refs.\,\cite{pla2012single, pla_high-fidelity_2013, morello_single-shot_2010}. A single electron transistor (SET) (depicted in cyan in \reffig{fig:setup}) is used to read out the spin states of the donor-bound electron~\cite{morello_single-shot_2010} or the nucleus~\cite{pla_high-fidelity_2013}. The device structure also comprises a broadband on-chip microwave magnetic antenna (brown on the right of \reffig{fig:setup}) to drive the spins via standard nuclear magnetic resonance (NMR) and electron spin resonance (ESR) techniques. Local gates are added in order to tune the tunnel coupling between the donor and the SET (`SET rate' gate, SR, red) and the coupling between the donor and interface quantum dot by laterally shifting the electron wavefunction~\cite{tosi_silicon_2017} (`right side' and `left side' gates, RS and LS, purple). The `fast donor' (FD) gate overlapping the implantation area is used as an electrical antenna (brown on the left of \reffig{fig:setup}) to apply an electrical microwave drive tone. 

The fabrication procedure of the donor qubit device follows the recipe outlined in Refs.\,\cite{pla2012single, mkadzik2022precision}. Here, we provide a short summary containing important implantation parameters and dimensions of the nanostructures. The natural silicon wafer that is used in this work contains a 900~nm thick isotopically enriched $^{28}$Si epitaxial layer (residual $^{29}$Si concentration of 730~ppm) on top of a lightly p-doped natural Si handle wafer. Using optical lithography and thermal diffusion of Phosphorus (Boron), n$^{+}$ (p) regions are defined on the sample. The n$^{+}$-type regions serve as electron reservoirs for the qubit device and are connected to aluminum Ohmic contacts. The p-doped regions are added to prevent leakage between Ohmic contacts. Using wet thermal oxidation, a 200~nm SiO$_{2}$ field oxide is grown on top of the substrate. The active device region is defined by etching a 30~$\mathrm{\mu}$m$\times$60~$\mathrm{\mu}$m area in the centre of the field oxide using HF acid which is subsequently covered by an 8~nm high-quality SiO$_{2}$ gate oxide in a dry thermal oxidation step. We define a 100~nm$\times$90~nm window (see red dashed box in \reffig{fig:setup}) in a 200~nm thick PMMA resist using electron-beam lithography (EBL), through which the $^{31}$P$^{+}$ ions are implanted at a 7$^{\circ}$ angle from the substrate norm. For this sample, we use an implantation energy of 10~keV and a fluence of $1.4\times10^{12}$~atoms/cm$^{2}$. According to Monte Carlo SRIM (Stopping and Range of Ions in Matter)~\cite{SRIM} simulations, approximately 40 donors are implanted within the window region $\approx$ 3.5 to 10.1~nm deep below the SiO$_{2}$/Si interface. To activate the donors and repair the implantation damage, we use a rapid thermal anneal at 1000 $^{\circ}$C for 5~s. To avoid potential leakage through the thin SiO$_{2}$ layer, we deposit an additional 3~nm Al$_{2}$O$_{3}$ layer via atomic layer deposition. The qubit device itself is defined in four EBL steps each including a thermal deposition of aluminum gates (with increasing thickness of 25~nm, 25~nm, 50~nm and 80~nm). After each deposition, the sample is exposed to a pure 100~mTorr oxygen gas for three minutes to oxidize an insulating Al$_{2}$O$_{3}$ layer between the gates. We connect all gate electrodes from all layers electrically to avoid electrostatic discharge damage (the shorts are broken after wire bonding using a diamond scriber). Finally, we anneal the qubit devices at 400 $^{\circ}$C in a forming gas (95\% N$_{2}$ / 5\% H$_{2}$) atmosphere for 15 minutes to passivate interface traps and repair EBL damage.

\textbf{Measurement setup}

The sample is wire-bonded to a gold-plated printed circuit board inside a copper enclosure. The enclosure is mounted onto a cold finger attached to the mixing chamber of a Bluefors BF-LD400 dilution refrigerator and is cooled down to $\approx$ 17~mK. The sample is placed in the centre of a superconducting magnet, operated in persistent mode at a magnetic field between $B_{0} \approx$ 0.9-1~T. The field is applied along the short-circuit termination of the magnetic (ESR, NMR) antenna, parallel to the surface of the chip and to the [110] Si crystal direction. 
A schematic of the experimental setup is shown in \reffig{fig:setup}. Static DC voltages from battery-powered Stanford Research Systems (SRS) SIM928 voltage sources are used to bias the metallic gate electrodes via homemade resistive voltage dividers at room temperature. The SET Top Gate (TG), Left Barrier (LB), Right Barrier (RB) and Right Side (RS) gates are connected via second-order low-pass RC filters with a 20~Hz cut-off. Gates used for loading/unloading the donor, i.e.~the Plunger (PL), Left Side (LS), SET Rate (SR) and Fast Donor (FD) gates, are filtered by a seventh-order low-pass LC filter with a 80~MHz cut-off frequency. All filters are thermally anchored at the mixing chamber stage of the dilution refrigerator. Baseband pulses from a LeCroy ArbStudio 1104 arbitrary waveform generator (AWG) are added via room temperature resistive voltage combiners to FD and SR. The ESR microwave (MW) signals are generated by an Agilent E8257D 50~GHz analog source, and we use an Agilent N5182B 6~GHz vector source to create radio-frequency (RF) signals for NMR control. Using a Marki Microwave DPX-1721 diplexer at room temperature, both high-frequency signals are routed to the magnetic antenna via a semi-rigid coaxial cable. A 10~dB attenuator is used for thermal anchoring at the 4~K stage. 

The MW signal for EDSR control is generated by a Keysight E8267D 44~GHz vector source. For single-sideband IQ modulation, RF pulses from the LeCroy ArbStudio 1104 AWG are fed to the in-phase (I) and quadrature-phase (Q) ports of the vector source. The high-frequency signal is attenuated by 10~dB at the 4~K stage and combined to the baseband control pulses at the mixing chamber using a Marki Microwave DPX-1721 diplexer. The combined signal is routed to the FD gate (electric/EDSR antenna). 

The SET current is amplified using a room temperature Femto DLPCA-200 transimpedance amplifier ($10^{7}$~VA$^{-1}$ gain, 50~kHz bandwidth) and a SRS SIM910 JFET amplifier (100~VV$^{-1}$ gain). The amplified signal is filtered using a SRS SIM965 analog 50~kHz low-pass Bessel filter and digitized by an AlazarTech ATS9440 PCI card. The above instruments are triggered by a SpinCore PulseBlasterESR-Pro. Software control of the measurement hardware and the generation of pulse sequences is done in Python using the QCoDeS~\cite{Qcodes} and SilQ~\cite{SilQ} framework.

\textbf{Electron spin readout and initialization}
\label{sec:eread}

The spin of the donor-bound electron is read out using energy-dependent tunneling into a cold electron reservoir. Due to the large electron Zeeman splitting in our experiment, this translates into a measurement of the $S_z$ spin eigenstates, \qd and \qu. The method is a modified version of the well-known Elzerman readout scheme~\cite{Elzerman2004, morello_singleshot_architecture}. The modification consists of using the island of the SET charge sensor as the cold charge reservoir that discriminates the spin eigenstates~\cite{morello_singleshot_architecture,morello_single-shot_2010}, rather than having separate charge sensors and charge reservoir.

\begin{figure*}[htbp]
\centering
\includegraphics{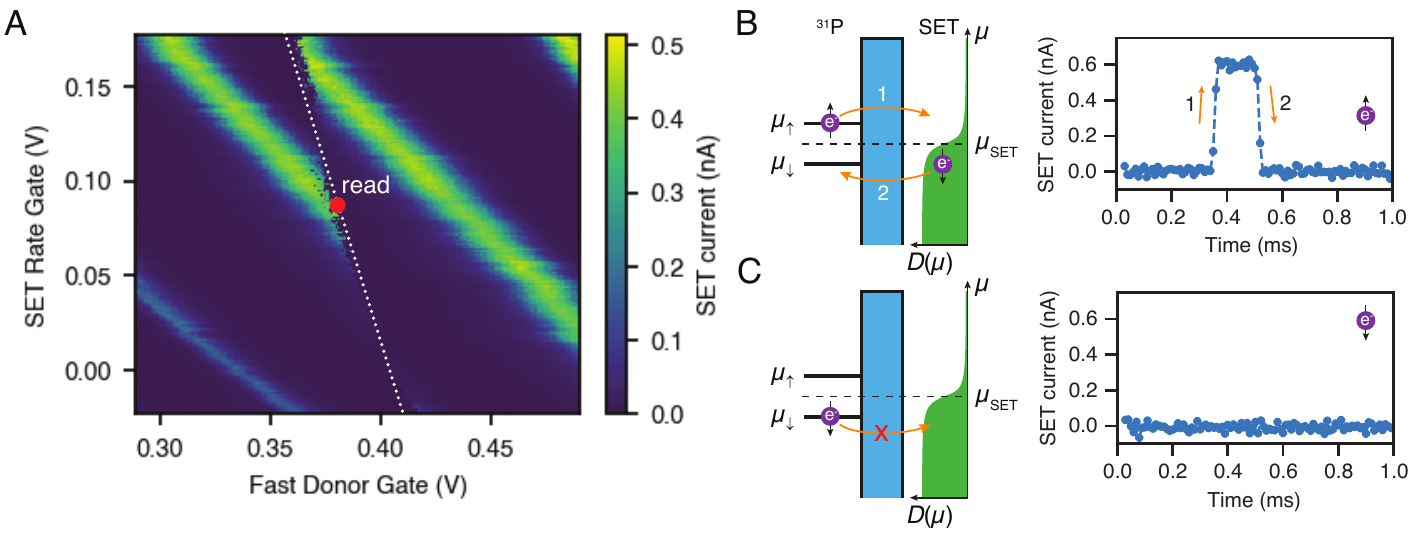}
\caption{\textbf{Electron spin-dependent tunneling.} (\textbf{A}) SET current as a function of two gate voltages. The pattern of Coulomb peaks (green) is broken (white dashed line) in the presence of a donor charge transition. The "read" position for the electron spin is indicated by a red dot, and corresponds to the point where the donor electrochemical potential equals that of the SET island. (\textbf{B,C}) Schematic depiction of the electron spin-dependent tunneling between the donor and the SET island (left panel) with the corresponding SET current traces (right panel). The $^{31}$P donor is tunnel-coupled to the SET island with a potential barrier between them shown in cyan. The Fermi-Dirac distribution for the density of the occupied states at the SET is shown in green.}
\label{SS-fig: electron-readout}
\end{figure*}

The white dotted line in \reffig{SS-fig: electron-readout}A highlights a donor charge transition. To perform electron spin readout and initialization, we tune the system into the so-called "read" spot (red dot in \reffig{SS-fig: electron-readout}A), where the electrochemical potentials of the SET island $\mu_\mathrm{SET}$ and the $^{31}$P donor $\mu_\mathrm{P}=(\mu_{\uparrow}+\mu_{\downarrow})/2$ are aligned, i.e.~$\mu_\mathrm{SET} = \mu_\mathrm{P}$.

In a static magnetic field $B_0$ ($\approx 1$~T), the Zeeman interaction splits the electrochemical potential of the donor into two energy levels, $\mu_{\downarrow}$ and $\mu_{\uparrow}$, for the two electron spin \qd and \qu states, respectively. In this case, $\mu_{\uparrow} > \mu_\mathrm{SET} > \mu_{\downarrow}$ and only the electron in the \qu state is energetically allowed to tunnel from the donor to the SET, as there are no available states below $\mu_\mathrm{SET}$ at the SET island (see \reffig{SS-fig: electron-readout}B). During this tunnel event (1 in the left panel of \reffig{SS-fig: electron-readout}B) the Coulomb blockade regime is lifted and we detect an increase in the SET current (right panel in \reffig{SS-fig: electron-readout}B). Since $\mu_{\downarrow}$ is the only level at the donor below $\mu_\mathrm{SET}$, only an electron in the \qd state can tunnel back from the SET to the donor (2 in the left panel of \reffig{SS-fig: electron-readout}B). In this case, the SET returns to the blockade regime and $I_\mathrm{SET} = 0$ again (right panel in \reffig{SS-fig: electron-readout}B). The obtained current spike, which we call a "blip", is then used as a spin-readout signal. It tells us that the electron at the donor was in the \qu state and is now initialized in the \qd state. If we do not detect any blips, it means the electron at the donor is in the \qd state and does not tunnel anywhere (\reffig{SS-fig: electron-readout}C). Therefore, this spin-dependent tunneling of the electron between the donor and the SET provides a single-shot readout and initialization of the electron spin into the \qd state~\cite{morello_single-shot_2010}.

The electron spin-up proportion $P(\uparrow)$ is then calculated by averaging over multiple repetitions. Further details of donor electron spin readout and initialization can be found in Refs.~\cite{morello_singleshot_architecture, morello_single-shot_2010,johnson2022beating}.

\textbf{Nuclear spin readout}
\label{sec:nread}

The donor-bound electron can be used as an ancilla qubit to read out the state of the nuclear spin via quantum non-demolition measurement with fidelities exceeding 99.99\% (see later section for more details).

The hyperfine interaction between the electron and the nucleus results in two electron resonance frequencies $f_\mathrm{ESR1} = \gamma_\mathrm{e}B_0 - \frac{A}{2}, f_\mathrm{ESR2} = \gamma_\mathrm{e}B_0 + \frac{A}{2}$, depending on the nuclear spin being $\Downarrow$ or $\Uparrow$~\cite{pla_high-fidelity_2013}. 
Starting from a \qd electron, an adiabatic ESR inversion pulse~\cite{laucht_high-fidelity_2014} at either ESR frequency results in a \qu electron if the nuclear spin was in the state corresponding to ESR frequency being probed. In other words, the ESR inversion constitutes a controlled-X (CX) logic operation on the electron, conditional on the state of the nucleus. We perform the electron inversion using an adiabatic frequency sweep across the resonance \cite{laucht_high-fidelity_2014} to be insensitive to small changes in the instantaneous resonance frequency.

Reading out the electron spin after an inversion pulse determines the nuclear state in a single shot (we call this readout a 'shot'). The fidelity of this readout process is the product of the single-shot electron readout fidelity, and the fidelity of inverting the electron via an adiabatic pulse.

Since $\gamma_\mathrm{e} B_0 \gg A$, the electron-nuclear hyperfine coupling is well approximated by $AS_z I_z$, meaning that the interaction commutes with the Hamiltonian of the nuclear spin. This is the quintessential requirement of a quantum non-demolition (QND) measurement~\cite{Braginsky1996, pla_high-fidelity_2013}. In practice, it means that the nuclear spin will be found again in the same eigenstate after the first shot. We can thus repeat it multiple times, i.e.~perform multiple measurement shots, to improve the nuclear readout fidelity. We calculate the electron spin-up proportion over all shots and determine the nuclear state by comparing the spin-up proportion to a threshold value (typically around 0.4-0.5).

For nuclear spin qubit manipulations that do not depend on the initial state (e.g.~Rabi drive), measuring the nuclear spin flip probability instead of the actual spin state is sufficient. For this, we read out the nuclear state $N_\mathrm{samples}$ times (typically $N_\mathrm{samples} \geq 20$) and calculate how many times the nuclear spin flipped ($N_\mathrm{flip}$) in two consecutive measurements. The flip probability is then determined as $P_\mathrm{flip} = N_\mathrm{flip} / (N_\mathrm{samples} - 1)$.

\textbf{Nuclear spin initialization}
\label{sec:ninit}

\begin{figure*}[htbp]
\centering
\includegraphics{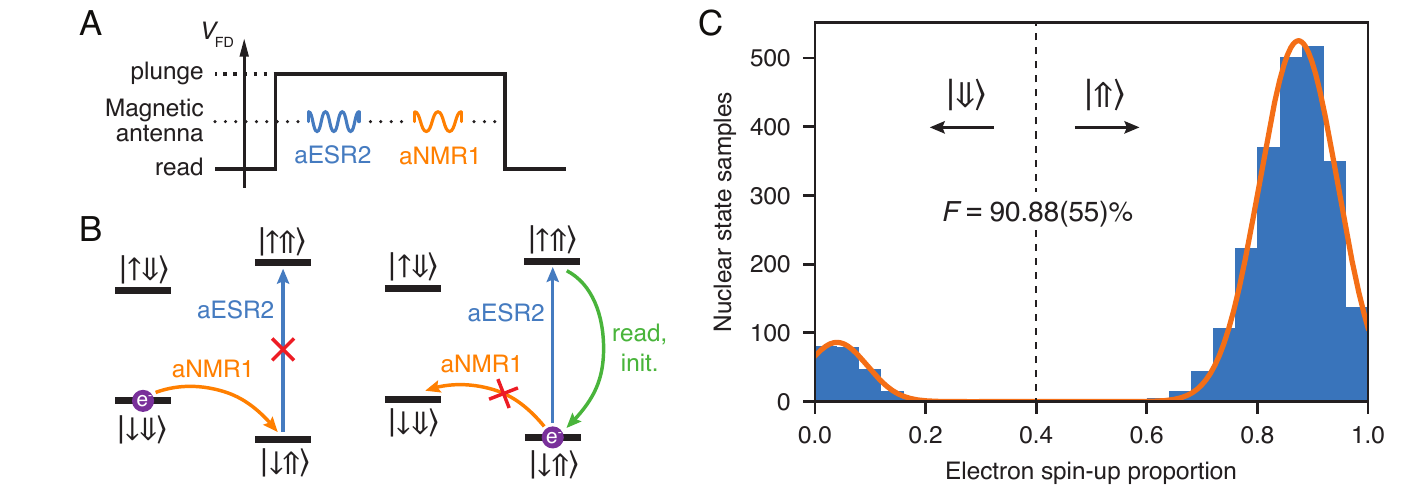}
\caption{\textbf{Nuclear spin \qU initialization fidelity.} (\textbf{A}) ENDOR sequence containing consecutive aESR2 and aNMR1 pulses. The sequence starts and ends in the "read" phase to initialize the electron into the \qd state. (\textbf{B}) Energy level diagrams for the $^{31}$P donor show nuclear spin initialization into the \qU state when we apply the ENDOR sequence in (\textbf{A}). With the electron initialized in the \qd state in the "read" phase, we consider two cases of initial nuclear states before the ENDOR sequence: \qD (left diagram), \qU (right diagram). (\textbf{C}) The histogram shows the final nuclear state after an ENDOR initialization sequence followed by nuclear spin readout with 25 shots. The nuclear spin is randomized after the 25 shots, to test the ENDOR sequence again. We assign to \qU the instances where the probability $P_{\uparrow}$ of finding the electron in the \qu state is above 0.4. Samples where $P_{\uparrow} < 0.4$ threshold correspond to the nuclear spin \qD state, i.e.~to the error of the ENDOR initialization. The histogram yields ENDOR initialization fidelity of $F = 90.88(55)\%$. The nuclear readout fidelity can be inferred by fitting the histogram peaks and calculating their overlap, yielding $F_{\rm read} > 99.99$\%. }
\label{fig:endor}
\end{figure*}

For benchmarking and quantum logic experiments, we need to initialize the flip-flop qubit into the \qdU state. We have seen that reading out the electron state also initializes it into the \qd state. To deterministically initialize the nucleus into the \qU state, and hence the flip-flop qubit into the \qdU, we make use of an electron-nuclear double resonance (ENDOR) sequence.
The ENDOR sequence comprises an adiabatic ESR2 (aESR2) pulse, an adiabatic NMR1 (aNMR1) pulse and an electron readout (see \reffig{fig:endor}A). We use adiabatic pulses that sweep around the actual resonance frequency and adjust the frequency range such that the pulses are insensitive to frequency deviations. On the one hand, if the system is in the \qdD state, the aESR2 pulse is off-resonant and the aNMR1 pulse flips the nuclear spin to the \qU state (left panel in \reffig{fig:endor}B). If, on the other hand, we already start in the \qdU state, the aESR2 pulse inverts the electron spin and the aNMR1 pulse is off-resonant (right panel in \reffig{fig:endor}B). After reading out and initializing the electron to the \qd state, the flip-flop qubit is initialized into the \qdU ground state. 

To quantify the nuclear spin-up probability $P(\Uparrow)$, we perform nuclear readout and determine the nuclear spin state by comparing the electron spin-up proportion (usually $\geq$ 20 shots) to a threshold value. By averaging over $N_\mathrm{samples} \geq 20$ repetitions, we derive $P(\Uparrow)$.

The fidelity of the ENDOR sequence is mostly limited by the errors of electron spin initialization in the \qd state since the fidelity of ESR/NMR inversion pulses is typically $> 98\%$. To measure these errors, we apply the ENDOR sequence and determine the nuclear state by measuring the electron spin-up proportion $P(\uparrow)$. We repeat this experiment 2500 times and apply multiple adiabatic NMR1 pulses after every 25 measurements to scramble the nuclear spin state. We determine the ENDOR fidelity to be $\mathcal{F} = 90.88(55)\%$ by calculating how many times we end up in the desired \qU state. The confidence interval shown in brackets represents the standard deviation, which is calculated by dividing 2500 experiments into 100 independent measurements (25 experiments each) of the nuclear spin-up state probability $P(\Uparrow)$. The histogram of nuclear readouts is shown in \reffig{fig:endor}C. The measured fidelity is in good agreement with the off-resonant nuclear spin-up probability of $P(\Uparrow) \approx 0.1$ that we see in EDSR spectrum scans.

\section*{Acknowledgments}
We acknowledge S. Asaad and M. Johnson for support with the measurement software SilQ, H. Firgau and V. Mourik for help during sample fabrication, and A. Ardavan, A. Heinrich, T. Proctor, and A. Kringh{\o}j for helpful conversations. 

\textbf{Funding:} The research was supported by the Australian Research Council (Grant no. CE170100012), the US Army Research Office (Contract no. W911NF-17-1-0200), and the NSW node of the Australian National Fabrication Facility. R.S. acknowledges support from Australian Government Research Training Program Scholarship. R.S. and I.F.d.F acknowledge support from the Sydney Quantum Academy. All statements of fact, opinion or conclusions contained herein are those of the authors and should not be construed as representing the official views or policies of the U.S. Army Research Office or the U.S. Government. The U.S. Government is authorized to reproduce and distribute reprints for Government purposes notwithstanding any copyright notation herein.

\textbf{Author contributions:} R.S., T.B. and I.F.d.F. designed and performed the measurements with A.M. supervision. R.S., T.B. and F.E.H. fabricated the device, with supervision from A.M. and A.S.D., on an isotopically enriched $^{28}$Si wafer supplied by K.M.I.. A.M.J, B.C.J. and D.N.J. designed and performed the ion implantation. B.J. performed the spatial triangulation of the donor under study. R.S., T.B. and A.M. wrote the manuscript, with input from all coauthors.

\textbf{Competing interests:} A.M. is a coauthor on the patent "Quantum processing apparatus and a method of operating a quantum processing apparatus" (US10528884B2, AU2016267410B2) which describes the invention and application of the flip-flop qubit. A.S.D. is a founder, equity holder, director and CEO of Diraq Pty Ltd.

\begin{widetext}
\textbf{Data and materials availability:} All the data supporting the contents of the manuscript, including raw measurement data and analysis scripts, can be downloaded from the following repository:

\noindent
https://datadryad.org/stash/share/4zGjpW4BkjAB7df8NgcsteNn-A1bD7aHaXtPRkLnE6Y
\end{widetext}

\DeclareRobustCommand{\citenumber}[1]{Ref.\,\citenum{#1}\xspace}
\graphicspath{{./images/}}
\newcommand{\cmmnt}[1]{\ignorespaces}

\maketitle
~
\newpage
~
\newpage

	\section*{Supplementary Material}
	
	\section{S1: Qubit Hamiltonian}
	The spin Hamiltonian of a $^{31}$P donor in silicon, expressed in frequency units, reads:
	\begin{equation}
		\hat{H} = \gamma_\mathrm{e} B_0 \hat{S}_z - \gamma_\mathrm{n} B_0 \hat{I}_z + A\mathbf{\hat{S}}\cdot \mathbf{\hat{I}}.
		\label{seq: Hamiltonian}
	\end{equation}
	$\mathbf{\hat{S}} = (\hat{S}_x, \hat{S}_y, \hat{S}_z)$ and $\mathbf{\hat{I}} = (\hat{I}_x, \hat{I}_y, \hat{I}_z)$ are the spin vector operators for the electron and nucleus, respectively.  The first (second) term represent the electron (nuclear) Zeeman interactions in a magnetic field $\vec{B}_0$ pointing along the $z$-axis and proportional to the electron (nuclear) gyromagnetic ratio $\gamma_\mathrm{e} \approx 27.97$~GHzT$^{-1}$ ($\gamma_\mathrm{n} \approx 17.25$~MHzT$^{-1}$). For convenience, we define the electron and nuclear gyromagnetic ratios as positive, hence the negative sign from the Zeeman interaction cancels out for the electron Zeeman term in Eq.\,\ref{seq: Hamiltonian}. The third term describes the isotropic Fermi contact hyperfine interaction, where $A = 117.53$~MHz for Phosphorus donors in bulk silicon. In the present device, where the donor is placed in strong electric fields and in proximity to metallic electrodes, $A=114.1$~MHz. At high magnetic fields ($B_0 \approx 1$~T) $\gamma_\mathrm{e} B_0 \gg A > 2\gamma_\mathrm{n} B_0$ and the eigenstates of this two-spin system are approximately the tensor product of electron-nuclear spin states: $\Ket{\downarrow, \uparrow} \bigotimes \Ket{\Uparrow, \Downarrow} \in \{ \qdU, \qdD, \quD, \quU \}$ (see Fig.\,1\,A). 
	
	The flip-flop qubit~\cite{tosi_silicon_2017} is encoded in the anti-parallel electron-nuclear spin states \qdU and \quD of the $^{31}$P donor (Fig.\,1\,A). As the qubit states have the same $z$-component of the total angular momentum, transitions between \qdU and \quD cannot be driven magnetically. However, the hyperfine interaction term in the Hamiltonian from Eq.\,\ref{seq: Hamiltonian} couples the flip-flop states, since its eigenstates are the singlet $\Ket{S} = (\qdU - \quD)/\sqrt{2}$ and triplet $\Ket{T_0} = (\qdU + \quD)/\sqrt{2}$ states (see Fig.\,1\,B). This can also be seen in the matrix form of the full Hamiltonian:
\begin{widetext}
	\begin{equation}
		\hat{H} = \frac{1}{2}
		\pmatrix{
			\gamma_{-} B_0 + A/2 & 0 & 0 & 0 \cr
			0 & \gamma_{+} B_0 - A/2 & A & 0 \cr
			0 & A & -\gamma_{+} B_0 - A/2 & 0 \cr
			0 & 0 & 0 & - \gamma_{-} B_0 + A/2 },
		\label{FF-eq: Hamiltonian matrix}
	\end{equation}
\end{widetext}
	where $\gamma_{\pm} = \gamma_\mathrm{e} \pm \gamma_\mathrm{n}$, and the columns of the matrix are ordered as the \quU, \quD, \qdU, and \qdD states. From this, the Hamiltonian in the truncated flip-flop qubit subspace is
\begin{widetext}
	\begin{equation}
		\hat{H}_\mathrm{ff} = \frac{1}{2}
		\pmatrix{
			\gamma_{+} B_0 - A/2 & A \cr
			A & -\gamma_{+} B_0 - A/2}
		=
		\frac{1}{2} (\gamma_{+} B_0 \hat{\sigma}_z + A \hat{\sigma}_x - \frac{A}{2} \hat{\mathbf{1}}),
		\label{FF-eq: flip-flop matrix}
	\end{equation}
\end{widetext} 
	
	Being orthogonal to the flip-flop basis ($\hat{\sigma}_x$-term in Eq.\,\ref{FF-eq: flip-flop matrix}), the hyperfine interaction $A$ can be used to perform electrically-driven spin resonance (EDSR) transition between the flip-flop states~\cite{laird2007hyperfine} (Fig.\,1\,A) by modulating the hyperfine coupling $A(t)$ at the flip-flop resonance frequency, which according to Eq.\,\ref{FF-eq: flip-flop matrix} is given by
	\begin{equation}
		\epsilon_\mathrm{ff} = \sqrt{(\gamma_\mathrm{+} B_0)^2 + A^2}.
		\label{FF-eq: EDSR frequency}
	\end{equation}
	
	As the hyperfine interaction is defined by the overlap of the electron wavefunction with the $^{31}$P nucleus, it can be modulated by displacing the electron from the donor nucleus using electric fields~\cite{park2009mapping, pica2014hyperfine, laucht_electrically_2015}. For an electric field dependent hyperfine interaction, we derive the Rabi frequency from the rotating wave approximation~\cite{laucht2016breaking} as:
	\begin{equation}
		2g^\mathrm{ff}_\mathrm{E} = f^{\rm ff}_\mathrm{Rabi} = \frac{1}{2} \frac{\partial A(E)}{\partial E} E_{\rm ac},
		\label{FF-eq: EDSR drive general}
	\end{equation}
	where $\partial A(E)/\partial E$ is the Stark shift of the hyperfine coupling and $E_{\rm ac}$ is the amplitude of the oscillating electric field. 
	
		\section*{S2: Capacitive triangulation of the donor location}
	The Phosphorus donors are implanted within a 100~nm$\times$90~nm region underneath the FD gate and in close proximity to the SET and magnetic antenna. This means that the exact location of the specific donor which we used as qubit is a priori unknown.
	
	To narrow down its possible location, we use a triangulation method based on comparing the capacitive couplings between the donor-bound electron and several electrostatic gates in the device~\cite{mohiyaddin2013noninvasive,asaad2020coherent}.
	
\begin{figure*}[htbp]
		\centering
		\includegraphics{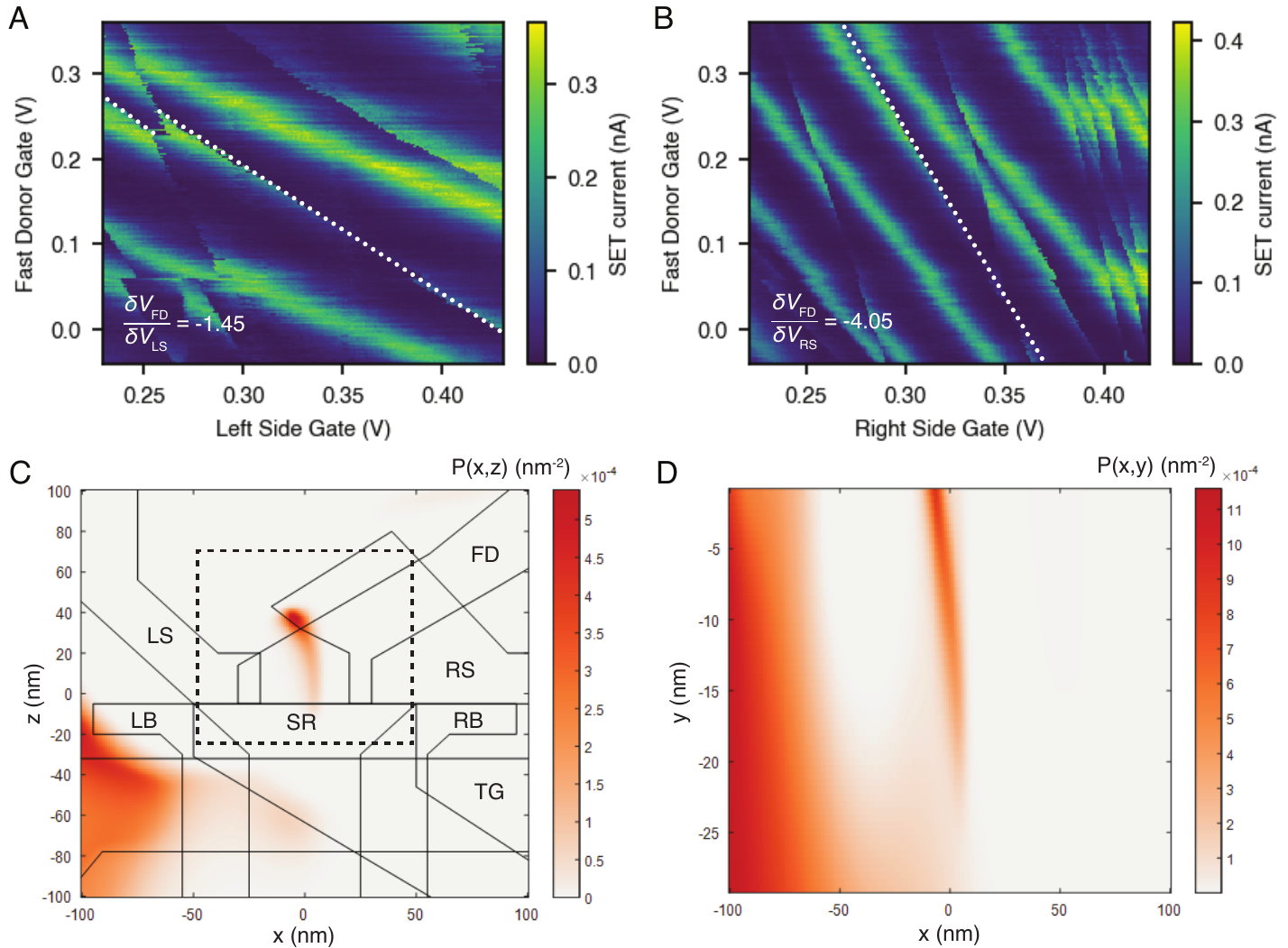}
		\caption{\textbf{Donor triangulation.} (\textbf{A-B}) Charge stability diagrams around the donor transition (white dotted line) using FD, LS and RS gates with the respective slopes of the transition shown in the bottom left corner. (\textbf{C-D}) Probability density distribution of the donor location in-plane of the substrate (\textbf{C}) and perpendicular to it (\textbf{D}). The distribution is calculated by comparing measured and simulated slopes of the donor transition (see Eq.\,\ref{CD-eq: probability estimate} for details). The dashed black region in (\textbf{C}) shows the implantation window in the device. We find a high probability at the tip of the RS gate.}
		\label{fig:triangulation}
	\end{figure*}

	We start by measuring charge stability diagrams around the donor transition using different gate electrodes (see \reffig{fig:triangulation}A-B as an example for the LS, RS and FD gates). The white dotted line is the so-called donor charge transition, where the electrochemical potentials of the donor and the SET are aligned. This means that the electrostatic potential $V(\vec{r}_0, V_\mathrm{FD}, V_\mathrm{RS}, V_\mathrm{LS}, \dots)$ at the donor location $\vec{r}_0$ is kept constant along the transition and the gate voltages in \reffig{fig:triangulation}A must satisfy the relation
	\begin{widetext}
		\begin{equation}
			\frac{\partial V(\vec{r}_0, V_\mathrm{FD}, V_\mathrm{LS}, V_\mathrm{RS}, \dots)}{\partial V_\mathrm{FD}} \delta V_\mathrm{FD} + \frac{\partial V(\vec{r}_0, V_\mathrm{FD}, V_\mathrm{LS}, V_\mathrm{RS}, \dots)}{\partial V_\mathrm{LS}} \delta V_\mathrm{LS} = 0,
			\label{CD-eq: donor potential}
		\end{equation}
	\end{widetext}
	where $\delta V_\mathrm{FD}$ and $\delta V_\mathrm{LS}$ are the respective gate voltage changes along the transition. This relation can be rewritten as
	\begin{equation}
		\frac{\delta V_\mathrm{FD}}{\delta V_\mathrm{LS}} = - \frac{\partial V(\vec{r}_0, V_\mathrm{FD}, V_\mathrm{LS}, \dots)}{\partial V_\mathrm{LS}} \bigg/ \frac{\partial V(\vec{r}_0, V_\mathrm{FD}, V_\mathrm{LS}, \dots)}{\partial V_\mathrm{FD}},
		\label{CD-eq: transition slope}
	\end{equation}
	where the ratio between the gate capacitances (left hand side) represents the slope of a transition, i.e.~$s_\mathrm{LS}$, in the charge stability diagram in \reffig{fig:triangulation}A. In the same way, we determine the slopes for the remaining gate combinations (see \reffig{fig:triangulation}B for RS and FD gates, the rest of the combinations are available in a public data repository). 
	
	Next, we perform simulations of the electrostatic potential landscape of the device using the COMSOL software package, to calculate the right hand side of Eq.\,\ref{CD-eq: transition slope} for positions $\vec{r}$ within the  200~nm$\times$200~nm area around the implantation window (see \reffig{fig:triangulation}C). In these simulations, we model the device according to our design layout and consider the 2DEG underneath the SET as a 1~nm thick metallic layer at the Si/SiO$_2$ interface, with lateral dimension reflecting those of the SET. 
	
	Having obtained the right hand side slopes $s^{\mathrm{sim}}_{g}(\vec{r})$ in Eq.\,\ref{CD-eq: transition slope} for all measured gates $g \in \mathrm{ \{LS, RS, \dots \} }$, we compare them to the experimental values $s_{g}$ using a least-squares estimate~\cite{sivia2006data, asaad2020coherent} as
	\begin{equation}
		P(\vec{r}) = N \exp \bigg[ - \frac{1}{2} \sum_{g \in \mathrm{ \{LS, RS, \dots \} } } \bigg( \frac{ s^{\mathrm{sim}}_{g}(\vec{r}) - s_{g}}{\sigma_g} \bigg)^2 \bigg],
		\label{CD-eq: probability estimate}
	\end{equation}
	which returns the maximum probability density $P(\vec{r_0})$ at each position $\vec{r_0}$, where the difference between simulated and measured slopes is minimal. $N$ is a normalization factor and $\sigma_g$ is the standard deviation error of the measured slope $s_{g}$ for gate $g$ which we define as
	\begin{equation}
		\sigma_g = 1 + \frac{1}{s^2_{g}}.
		\label{CD-eq: slope error}
	\end{equation}
	In this way we give more weight to the gates that have larger slopes, i.e.~stronger capacitive coupling to the donor, since their effect can be estimated more reliably. The gates with smaller slopes are less reliable as they are strongly screened by the nearby gates and the 2DEG, and hence provide less accurate information about the donor position.
	
	As seen in \reffig{fig:triangulation}C, the capacitive model yields two regions where the $^{31}$P donor could be located: at the bottom-left corner of the device, i.e.~under the TG and near the LB gates, and underneath the tip of the RS gate. It is very unlikely that the donor is in the first region, since the capacitive coupling of the TG and LB gates to the donor-bound electron is small and this region is outside of the implantation window (black dashed region in \reffig{fig:triangulation}C). We thus ignore this region and consider the donor to be under the tip of the RS gate, slightly behind the FD gate (electric antenna). This location may explain the linear increase of the hyperfine interaction with the FD gate voltage (Fig.\,3\,B of the main text) instead of the expected decrease, since we increase the electron wavefunction overlap with the donor nucleus when applying a positive voltage on the FD gate~\cite{laucht_electrically_2015}. Further possible explanations for an increase in hyperfine coupling with positive FD gate voltage might involve a shallow ($<$ 3.2~nm) depth of the donor under the Si/SiO$_2$ interface. This would cause a strong electron wavefunction distortion by the interface barrier~\cite{mohiyaddin2014designing}. Strain at the donor location from the difference in the thermal expansion coefficients of the aluminum gates and the silicon substrate is another mechanism that can lead to a distortion of the electron wavefunction and a positive hyperfine tunability~\cite{dreher2011electroelastic, laucht_electrically_2015, mansir2018linear, asaad2020coherent}. To exactly identify the contribution from each mechanism would require additional investigation, for example, by using atomistic tight-binding simulations of the hyperfine coupling that include electric fields and strain in the vicinity of the expected donor location from the additional COMSOL simulations \cite{tosi_silicon_2017, laucht_electrically_2015, mohiyaddin2013noninvasive}. The capacitive triangulation method also provides information about the vertical (donor depth, $y$-axis) location of the donor (see \reffig{fig:triangulation}D). However, due to the planar gate layout of the donor device, the sensitivity (and hence the precision) of this method in the $y$-direction is low, which is why we find a large range of high probability density spanning almost 20~nm depth in \reffig{fig:triangulation}D.

	\section*{S3: EDSR drive amplitude calibration}
	
	\begin{figure*}[htbp]
		\centering
		\includegraphics{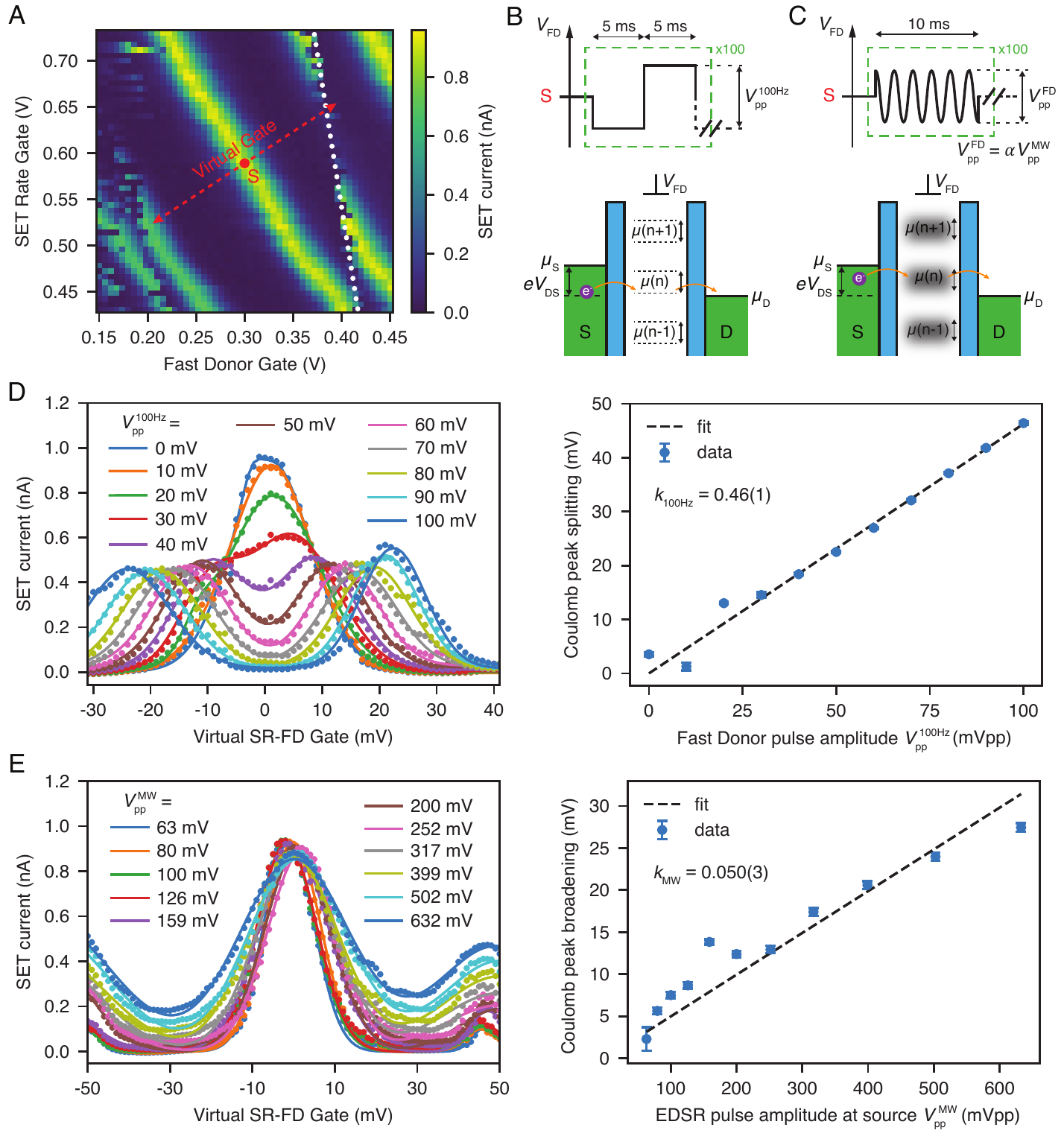}
		\caption{\textbf{EDSR drive amplitude calibration.} (\textbf{A}) The charge stability diagram shows the SET Coulomb peak next to the donor charge transition (white dotted line) used for the calibration procedure outlined in the text. The SET current peak is measured along the red dashed line. (\textbf{B}) A 100~Hz square wave pulse periodically splits the electrochemical potential of the SET island in two levels that are occupied half the time. (\textbf{C}) A high-frequency EDSR tone applied to the FD gate smears out and broadens the electrochemical potential. (\textbf{D}) SET current as a function of the amplitude of the square wave depicted in \textbf{B}. The current peak splits in two upon increasing the amplitude, since we average the SET current over many periods of the square wave. We fit two Gaussians to the Coulomb peak doublets to extract the peak splitting as a function of the amplitude of the 100~Hz square wave. (\textbf{E}) The MW tone (panel \textbf{C}) results in a broadening rather than a splitting of the SET current peak. We fit three Gaussians to the Coulomb peaks to determine the dependence of the central Coulomb peak broadening on the amplitude of the EDSR pulse at the MW source. Comparing the effects of the 100~Hz square wave and the MW tone allows to infer the drive amplitude $V_{\rm pp}^{\rm MW}$, since the low-frequency square wave amplitude $V_{\rm pp}^{\rm 100Hz}$ is accurately known.}
		\label{fig:calibration}
	\end{figure*}
	
	Assuming the electric field at the donor is proportional to the voltage applied to the FD gate, we can rewrite the equation from the main text as
	\begin{equation}
		f_{\rm Rabi}^{\rm ff} = \frac{1}{2} \frac{\partial A(E)}{\partial E}E_{\rm ac} =
		\frac{1}{2} \frac{\partial A(V_\mathrm{FD})}{\partial V_\mathrm{FD}} \Delta V_\mathrm{FD},
		\label{CD-eq: EDSR drive vs FD}
	\end{equation}
	where $\frac{\partial A(V_\mathrm{FD})}{\partial V_\mathrm{FD}}$ is the hyperfine tunability with the applied voltage to the FD gate and $\Delta V_\mathrm{FD}$ is the amplitude of voltage oscillations on the FD gate during the EDSR pulse, which we will call the EDSR drive amplitude. The amplitude of these oscillations can be further presented as $\Delta V_\mathrm{FD} = \alpha V_\mathrm{MW}$, where $V_\mathrm{MW}$ is the peak-to-peak amplitude of EDSR pulse at the MW source and $\alpha$ is a coefficient representing the attenuation of the EDSR drive amplitude between the MW source and the tip of the electric antenna. The attenuation of the EDSR drive amplitude occurs in the coaxial cables, including a 10~dB attenuator at the 4~K stage and a diplexer at the mixing chamber, at the PCB and the electric antenna itself. As a result, Eq.\,\ref{CD-eq: EDSR drive vs FD} becomes
	\begin{equation}
		f^\mathrm{ff}_\mathrm{Rabi} = \frac{1}{2} \frac{\partial A(V_\mathrm{FD})}{\partial V_\mathrm{FD}} \alpha V_\mathrm{MW}.
		\label{CD-eq: EDSR drive vs MW}
	\end{equation}
	Comparing the slopes of the measured Rabi frequencies $f^\mathrm{ff}_\mathrm{R}$ as a function of microwave power (Fig.\,3\,A) to the DC Stark shift of the hyperfine coupling (Fig.\,3\,B) yields the following conversion factor between DC and AC signals (or line attenuation)
	\begin{equation}
		\alpha_\mathrm{EDSR} = 2\frac{\partial f^\mathrm{ff}_\mathrm{R}}{\partial V_\mathrm{FD}} \bigg/ \frac{\partial A(V_\mathrm{FD})}{\partial V_\mathrm{FD}} = 0.125(7) \equiv -18.1(5)~\mathrm{dB},
		\label{CD-eq: EDSR attenuation}
	\end{equation}
	where the confidence interval in the brackets represents a standard deviation. 
	
	We verify this result by an independent calibration of the MW to low-frequency conversion factor based on comparing the effect of a 100~Hz square wave and a 28~GHz MW sinusoid on the broadening of the SET Coulomb peak (along the red dashed line in \reffig{fig:calibration}A).
	Supplementary Figures \ref{fig:calibration}\,B,D show the splitting of the Coulomb peak as a function of the peak-to-peak amplitude $V^\mathrm{100Hz}_\mathrm{pp}$ of the 100~Hz square wave, where we average the SET current signal for 1~s at every gate voltage point. We fit the data to two Gaussian functions and determine the Coulomb peak splitting $\Delta V_\mathrm{100Hz}$ as the distance between their mean values. We obtain a linear dependence of $\Delta V_\mathrm{100Hz}$ on the amplitude $V^\mathrm{100Hz}_\mathrm{pp}$ of the FD pulse (see \reffig{fig:calibration}D), described by 
	\begin{equation}
		\Delta V_\mathrm{100Hz} = k_\mathrm{100Hz} V^\mathrm{100Hz}_\mathrm{pp},
		\label{CD-eq: splitting vs amplitude}
	\end{equation}
	where $k_\mathrm{100Hz} = 0.46(1)$. The splitting $\Delta V_\mathrm{100Hz}$ is almost half of $V^\mathrm{100Hz}_\mathrm{pp}$, since we apply the square wave pulse to FD only, but scan across the Coulomb peak in the SR direction as well (\reffig{fig:calibration}A) by means of a `virtual gate'.
	
	The Coulomb peak broadening due to a microwave tone (applied at the flip-flop resonance frequency in order to calibrate at the frequency of interest, although the spin dynamics has no bearing on the experiment) is shown in \reffig{fig:calibration}C,E. We extract the broadening as the FWHM $\Delta V$ of the middle peak by fitting the data to three Gaussian peaks. The excess broadening due to the EDSR tone is then calculated as $\Delta V_\mathrm{MW} = \sqrt{\Delta V^2 - \Delta V_\mathrm{ref}^2}$, by subtracting a reference value recorded without the drive tone applied~\cite{dehollain2012nanoscale}. The linear dependence of the Coulomb peak broadening on the EDSR pulse amplitude $V^\mathrm{MW}_\mathrm{pp}$ at the MW source, shown in \reffig{fig:calibration}E, is described by 
	
	\begin{equation}
		\Delta V_\mathrm{MW} = k_\mathrm{MW} V^\mathrm{MW}_\mathrm{pp},
		\label{CD-eq: broadening vs amplitude}
	\end{equation} 
	
	where $k_\mathrm{MW} = 0.050(3)$. Deviations of the data from the linear dependence in \reffig{fig:calibration}E are mainly attributed to thermal heating of the device during high-amplitude EDSR pulses, which contributes to the broadening of the Coulomb peak and leads to an increase of the offset of the current trace, which is currently poorly understood.
	
	Equating Eqs.\,\ref{CD-eq: splitting vs amplitude} and \ref{CD-eq: broadening vs amplitude} and substituting $V^\mathrm{100Hz}_\mathrm{pp} = \tilde{\alpha} V^\mathrm{MW}_\mathrm{pp}$ we derive the conversion factor between MW and 100~Hz signals to:
	\begin{equation}
		\tilde{\alpha} = \frac{k_\mathrm{MW}}{k_\mathrm{100Hz}} = 0.107(7) \equiv -19.4(5)~\mathrm{dB}.
		\label{CD-eq: SET attenuation}
	\end{equation}
	
	The independently determined conversion factor agrees well with the one extracted from Rabi and hyperfine measurements. The slight discrepancy of 1.3~dB can be explained by a change in lever arm of the FD gate to the SET island and the donor itself, i.e.~the ratio  between DC and AC electric field can be different at both locations. Thermal broadening and rectification effects due to high power electric signals and changes in the  resonance frequency due to spin flips of nearby $^{29}$Si nuclei in combination with a frequency-dependent line attenuation can also influence the estimate of the conversion factors.
	
	The good numerical agreement between the two estimates affirms that the Rabi drive strength is given by the Stark shift of the hyperfine coupling and that hyperfine-mediated EDSR is the driving mechanism for the flip-flop qubit.

	\section*{S4: Electron and flip-flop relaxation}
	\label{sec:T1}
	Early experiments on $^{31}$P donor ensembles in bulk silicon have shown an extremely slow relaxation process within the flip-flop qubit subspace ($\ket{\uparrow\Downarrow}\rightarrow\ket{\downarrow\Uparrow}$), $T_{\rm 1ff} \approx 5$~hours~\cite{feher1959electron} (note that the flip-flop process was labeled $T_x$ in the old literature). 
	
	Here, however, we deal with a near-surface donor, in close proximity to an oxide interface and several metallic gates. In the limit where the donor-bound electron hybridizes with an interface quantum dot, the flip-flop relaxation time $T_{\rm 1ff}$ can be reduced significantly~\cite{tosi_silicon_2017,boross2016valley}. We thus set out to measure both the electron, $T_{\rm 1e}$, and the flip-flop relaxation times directly.  
	
	We first determine the total electron spin relaxation time 
	\begin{equation}
		\frac{1}{T_1} = \frac{1}{T_{\rm 1e}} + \frac{1}{T_{\rm 1ff}}.
		\label{CD-eq: T1 relaxation}
	\end{equation}
	
\begin{figure*}[htbp]
		\centering
		\includegraphics{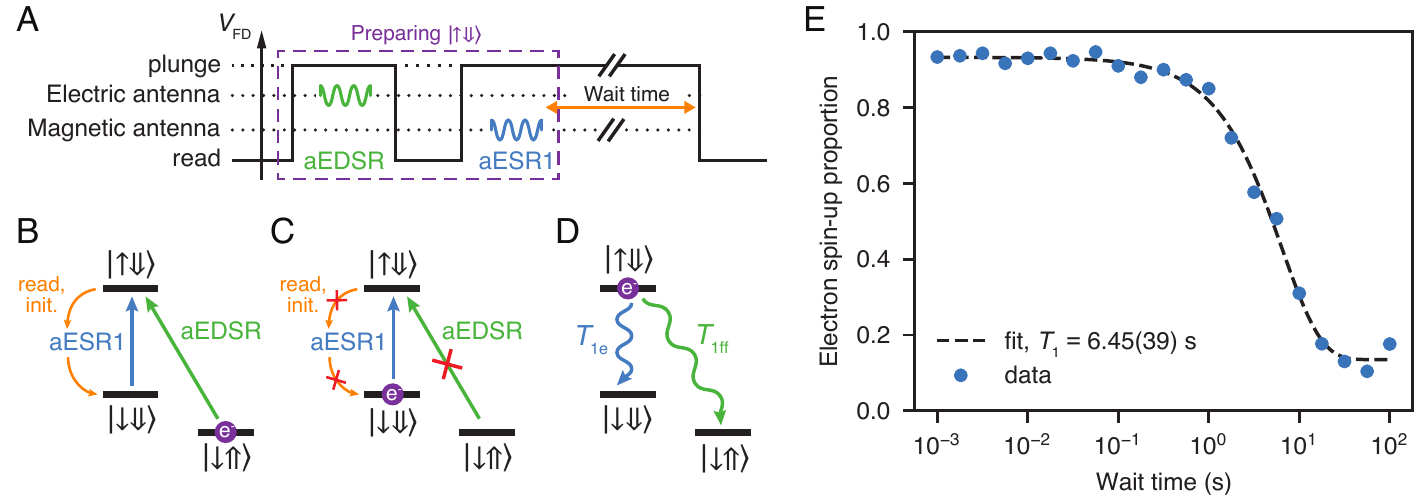}
		\caption{\textbf{Electron spin relaxation.} (\textbf{A}) A combination of aEDSR, electron initialization and aESR1 pulses is used to prepare the  excited \quD flip-flop state. (\textbf{B}) If the system is in \qdU, the pulse sequence prepares \quD via \qdU $\rightarrow$ \quD $\rightarrow$ \qdD  $\rightarrow$ \quD. (\textbf{C}) If the electron is in \qdD initially, the aEDSR pulse is off-resonant and we flip the electron spin to \quD via the aESR1 pulse. (\textbf{D}) The excited flip-flop \quD state can relax via two processes: electron spin relaxation into the \qdD state (blue, $T_{\rm 1e}$) and hyperfine-mediated flip-flop relaxation into the \qdU state (green, $T_{\rm 1ff}$). (\textbf{E}) The electron spin-up proportion as a function of the wait time in (\textbf{A}) shows the electron spin decay out of the \qu state. By fitting an exponential function (see text) to the data, we determine the relaxation time $T_{1} = 6.45(39)$~s, encompassing both the $T_{\rm 1e}$ and the $T_{\rm 1ff}$ processes.}
		\label{CD-fig: T1 relaxation}
	\end{figure*}

	We use a combination of aEDSR, electron read and aESR1 pulses to initialize the system in the excited \quD flip-flop state. The pulse sequence is shown in \reffig{CD-fig: T1 relaxation}A. We determine $T_1$ by measuring the electron decay from the \quD state which we fit to an exponential function $P\exp(-t / T_1) + P_\mathrm{offset}$. Here, $P_\mathrm{offset}$ is the electron spin-up proportion when the electron spin is fully decayed into the spin \qd state, and $P$ is the measurement contrast between the spin \qu and \qd states. We find $T_1 = 6.45(39)$~s, which is comparable to the typical electron spin relaxation times ($T_1 \approx T_{\rm 1e}$) in $^{31}$P donor qubit devices~\cite{tenberg2019electron}. According to Eq.\,\ref{CD-eq: T1 relaxation}, this suggests that $T_{\rm 1ff} > T_{\rm 1e}$ and the flip-flop relaxation process gives a negligible contribution to the total electron spin relaxation rate. 
	
\begin{figure*}[htbp]
		\centering
		\includegraphics{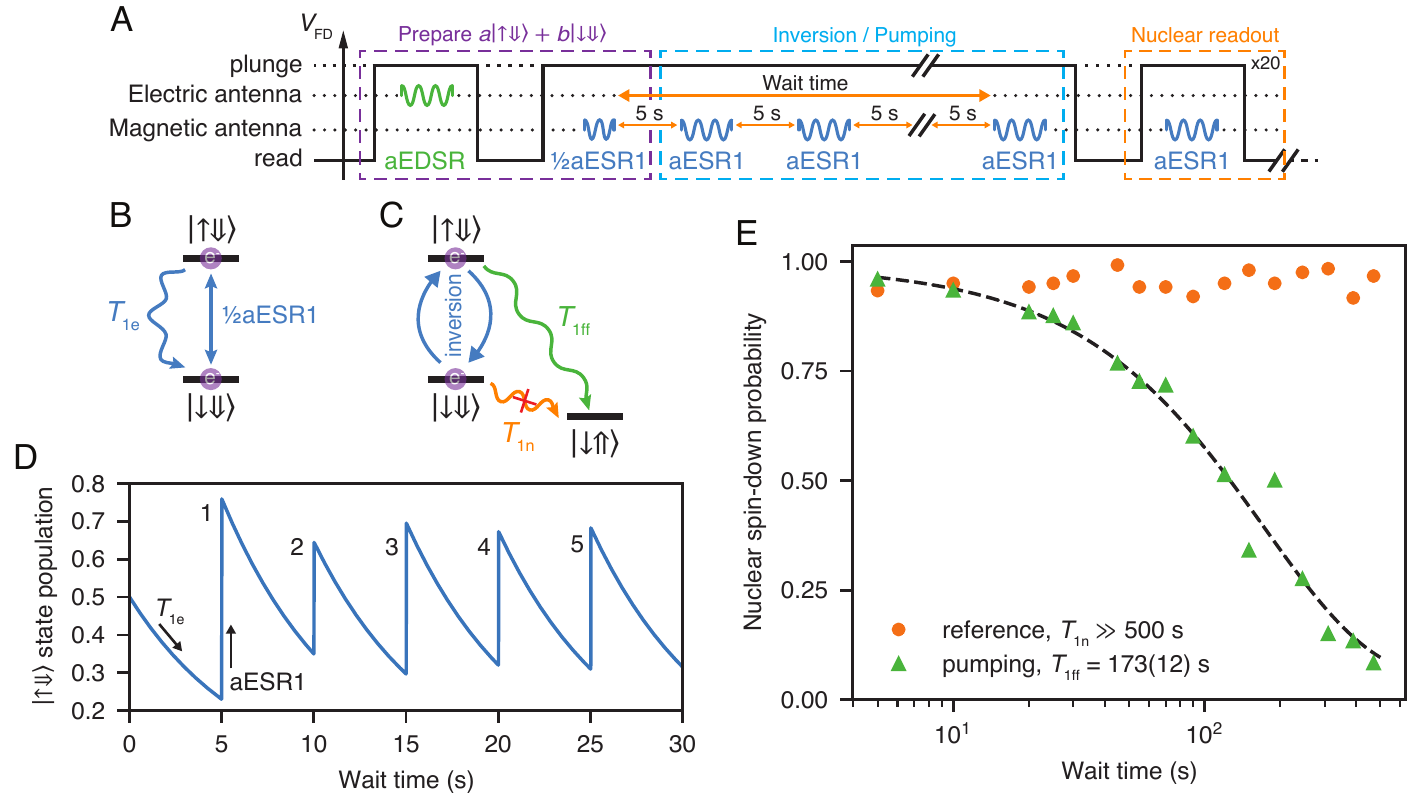}
		\caption{\textbf{Flip-flop relaxation.} (\textbf{A}) The pulse sequence used to measure the flip-flop relaxation time $T_{\rm 1ff}$ first prepares a superposition state of \qdD and \quD (purple part). Repeatedly applying aESR1 pulses in a 5~s interval ensures a non-zero population of \quD (blue part). The nuclear decay is measured by measuring the nuclear spin-down probability (orange part). (\textbf{B}) We prepare the donor in a superposition $a \qdD + b \quD$ state with $|a|^2 \approx |b|^2 \approx 0.5$. (\textbf{C}) By repeatedly inverting the population between \quD and \qdD using aESR1 pulses, we keep the excited flip-flop state populated and are able to measure the relaxation process $T_{\rm 1ff}$ via nuclear decay into the \qdU state. There is no direct nuclear spin relaxation $T_{\rm 1n}$ since the nucleus is decoupled from the environment. (\textbf{D}) The calculated population of the \quD state for the first 30~s of the wait time in the pulse sequence in (\textbf{A}). Because of a 5~s-delay between the aESR1 inversion pulses, the population decays due to $T_{\rm 1e}$ relaxation process. We consider $T_{\rm 1e} \approx 6.45$~s, the inversion fidelity of the aESR1 pulses is $98\%$, and neglect the population decay due to the flip-flop relaxation process in the first 30~s. (\textbf{E}) The nuclear spin-down probability dependence on the duration of the wait time in the sequence in (\textbf{A}) (green triangles). From the exponential fit (see text), the flip-flop relaxation time is estimated to be $T_{\rm 1ff} = 173(12)$~s. To demonstrate the absence of the nuclear spin relaxation $T_{\rm 1n}$ process, we perform a reference measurement without the aESR1 inversion pulses (orange dots).}
		\label{CD-fig: T1 EDSR relaxation}
	\end{figure*}

	Therefore, measuring $T_{\rm 1ff}$ requires a method to prevent the electron spin relaxation $\quD \rightarrow \qdD$ from bypassing the flip-flop process. We counteract the $T_{\rm 1e}$ process by applying the pulse sequence shown in \reffig{CD-fig: T1 EDSR relaxation}A. We first prepare \qdD  using an aEDSR pulse and an electron initialization pulse. We then create a superposition state $\ket{\psi_\mathrm{s}} = a \qdD + b \quD$ with equal population (i.e. ($|a|^2 \approx |b|^2 \approx 0.5$)) of the electron in the \quD and \qdD states by applying a $1/2$aESR1 pulse, i.e. a semi-adiabatic frequency sweep, with rate calibrated to yield an $\approx 50$\% probability of exciting the electron from the \qdD to the \quD state. Then, by repeatedly applying full-inversion aESR1 pulses (\reffig{CD-fig: T1 EDSR relaxation}B), we periodically reverse the effect of the $\quD \rightarrow \qdD$ relaxation channel, effectively saturating the ESR1 transition. Due to memory limitations of the AWG  we are only able to apply inversion pulses  every 5~s (see \reffig{CD-fig: T1 EDSR relaxation}A). Numerical simulation shows that this sequence leads to oscillations of the \quD state population between $\approx 0.23$ and $\approx 0.76$ with a mean value of $\approx 0.46$ (see \reffig{CD-fig: T1 EDSR relaxation}D), calculated by considering a $98\%$ fidelity of the inversion pulses and the electron spin relaxation time $T_{\rm 1e} \approx 6.45$~s. 
	
	We then observe the flip-flop relaxation (\reffig{CD-fig: T1 EDSR relaxation}C), i.e.~the $\quD \rightarrow \qdU$ process, by measuring the nuclear \qD state probability as a function of the wait time between the $1/2$aESR1 pulse and the last aESR1 pulse in the sequence (see \reffig{CD-fig: T1 EDSR relaxation}A and E). By fitting an exponential function to the data in \reffig{CD-fig: T1 EDSR relaxation}E we find $T_{\rm 1ff} = 173(12)$~s. As predicted earlier, the flip-flop relaxation time $T_{\rm 1ff} \gg T_{\rm 1e}$ and is indeed not a limiting factor in the flip-flop qubit operations discussed in this work. To rule out any nuclear relaxation process, we also perform a reference measurement, where we omit the inversion pulses after the $1/2$aESR1 pulse. In this case, the system simply relaxes into the \qdD state and we observe no further leakage out of that state, i.e.~no $\qdD \rightarrow \qdU$ process, yielding $T_{\rm 1n} \gg 500$~s, see \reffig{CD-fig: T1 EDSR relaxation}C,\,E. 
	
	\section*{S5: Pulse induced resonance shifts}
	\label{sec:stark}
	In this section, we investigate the physical origin of the difference in coherence times between the donor-bound electron qubit and the flip-flop qubit. We find that applying an electric drive simultaneously with the magnetic drive decreases the electron Rabi and Hahn echo coherence times, but not the Ramsey coherence time. For strong electric drive tones, we observe a frequency shift of the ESR, NMR and EDSR resonance frequencies depending on the duration and amplitude of the electric tone. The physical origin of those pulse-induced resonance frequency shifts (PIRS)~\cite{freer2017single} is not yet understood. Below we discuss further data on the present device, which clearly unveil the presence of PIRS and highlight some of its empirical features.
	
	\subsubsection*{Coherence measurements}
	
	\begin{figure*}[htbp]
		\centering
		\includegraphics{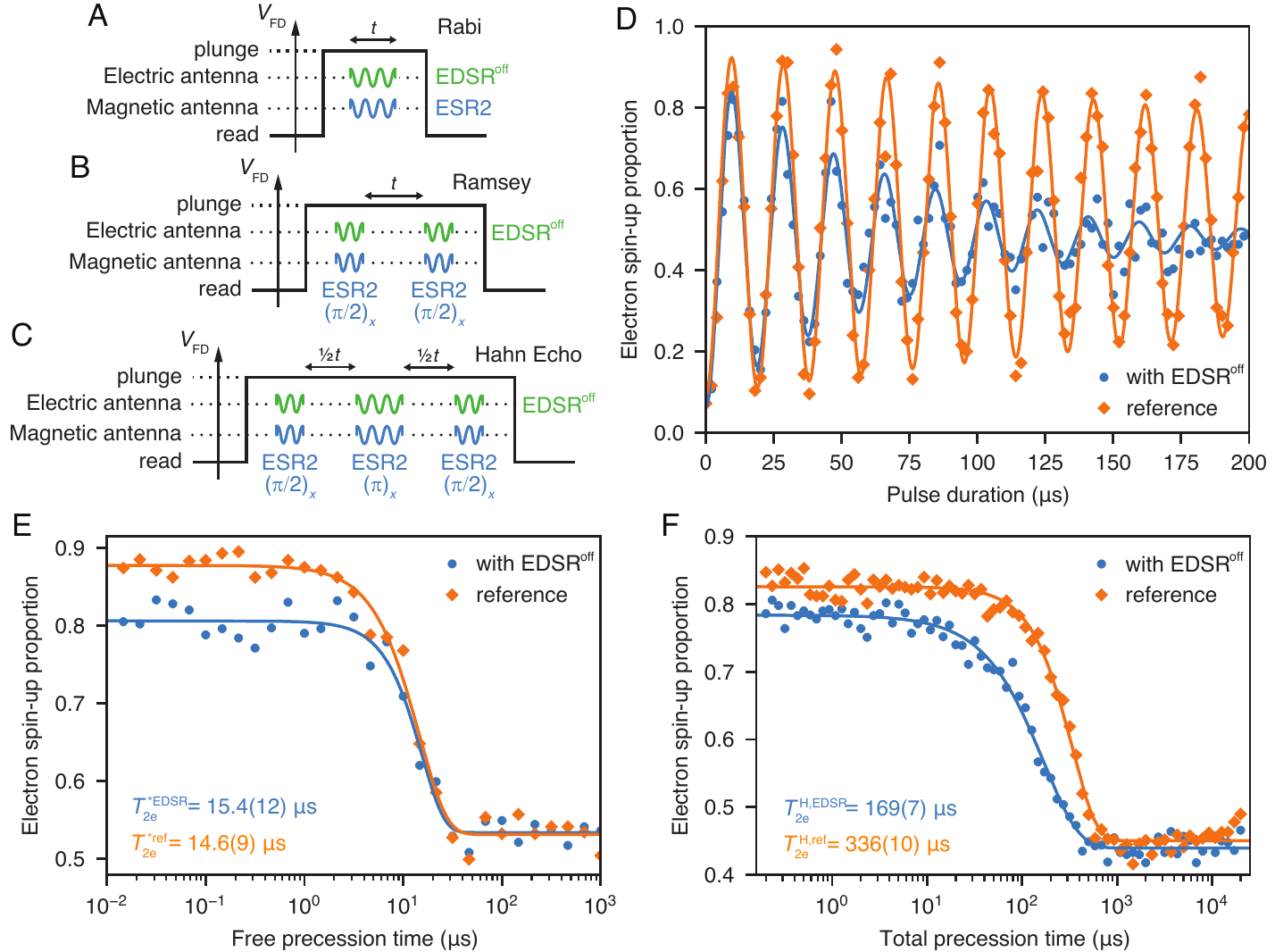}
		\caption{\textbf{Electron spin coherence.} (\textbf{A-C}) Pulse sequences used to measure the electron Rabi (\textbf{A}), Ramsey (\textbf{B}), and Hahn echo (\textbf{C}) decay times while simultaneously applying EDSR$^\mathrm{off}$ pulses only during the ESR pulses. (\textbf{D}) Electron spin Rabi oscillations obtained with (blue dots) and without (orange diamonds) the EDSR$^\mathrm{off}$ tone. The data are fitted with an exponentially decaying sinusoid $P \mathrm{exp}(-t / \tau^{\mathrm{R}}_\mathrm{e})\sin(2\pi f^{\mathrm{R}}_\mathrm{e} t + \phi) + P_{\infty}$ revealing that $\tau_{\rm e}^{\rm R}$ decreases from 460(73)~$\mu$s to 69(6)~$\mu$s in the presence of the EDSR$^\mathrm{off}$ tone. (\textbf{E-F}) Electron Ramsey (\textbf{E}) and Hahn echo (\textbf{F}) decays measured with (blue dots) and without (orange diamonds) the EDSR$^\mathrm{off}$ pulses. The data are fitted with an exponential decay function $P \exp(- (t / T_{\rm 2e})^{\beta} ) + P_{\infty}$ revealing the electron dephasing time $T^\mathrm{*}_{\rm 2e} = 14.6(9)~\mu$s without and $15.4(12)~\mu$s with the EDSR$^\mathrm{off}$ pulses. The exponents $\beta^\mathrm{*}$ of the decay are $1.58(17)$ and $1.91(33)$, respectively. The Hahn echo time $T^\mathrm{H}_{\rm 2e}$ decreases from $336(10)~\mu$s to $169(7)~\mu$s when applying the off-resonance EDSR$^\mathrm{off}$ pulses. The exponent $\beta^\mathrm{H}$ of the Hahn echo decay also decreases from $1.7(1)$ to $1.17(7)$, indicating a potential change in the spectrum of the noise.}
		\label{fig:ESR coherence}
	\end{figure*}
	
	The coherence times of the flip-flop qubit (summarized in Fig.\,4\,C) are consistently shorter than those of the electron spin qubit, including a shorter decay time of the driven Rabi oscillations, 100~$\mu$s for the flip-flop qubit compared to the $\geq 400~\mu$s for the electron. 
	
	Shorter flip-flop coherence times are to be expected if the system is operated in the large electric dipole regime, where the electron is significantly displaced from the donor nucleus, towards the Si/SiO$_2$ interface. That regime increases the system exposure to charge noise, although theory models predict the existence of a second-order clock transition where coherence may be protected~\cite{tosi_silicon_2017}. Here, however, we operate the donor qubit in a near-bulk regime (see S9), so one might expect the flip-flop decoherence to be dominated by the electron spin effects alone. On the other hand, the application of strong microwave electric fields may introduce effects that are not generally present when driving an electron spin qubit with oscillating magnetic fields.
	
	To investigate these effects, we measure the electron spin Rabi, Ramsey and Hahn echo times while applying an electric (EDSR$^\mathrm{off}$) tone simultaneously with the magnetic drive used for ESR (\reffig{fig:ESR coherence}). We choose an EDSR$^\mathrm{off}$ tone with half the amplitude of those used for the flip-flop coherence measurements in the main text, and offset in frequency by 5~MHz in order to avoid inducing any resonant spin transitions. For comparison, we also perform reference measurements without the additional EDSR$^\mathrm{off}$ tone. We fit the data (\reffig{fig:ESR coherence}D-F) with a damped sinusoid $P\mathrm{exp}(-t / \tau^{\rm R})\sin(2\pi f^{\mathrm{R}} t + \phi)+ P_{\infty}$ (Rabi) and an exponential decay $P\mathrm{exp}(-(\tau / T_2)^{\beta}) + P_{\infty}$ (Ramsey and Hahn echo). Here, $P$ is the amplitude, $P_{\infty}$ is the offset, $f^{\mathrm{R}}$ is the frequency, $\phi$ is the initial phase, $t$ is the duration of the Rabi oscillations and $\tau^{\rm R}$ is the Rabi decay time. In Ramsey and Hahn echo decay equation, $\tau$ is the total precession time, $T_2$ is the decay (coherence) time and $\beta$ is the exponent of the decay.
	
	In \reffig{fig:ESR coherence}D, we see that the electron Rabi decay time $\tau_{\rm e}^{\rm R}$ decreases from 460(73)~$\mu$s to 69(6)~$\mu$s when the electric drive is applied simultaneously. The Ramsey coherence time in \reffig{fig:ESR coherence}E is not affected, but we measure a decrease in readout contrast once we apply the EDSR$^\mathrm{off}$ tone. In \reffig{fig:ESR coherence}F we show the Hahn echo measurement. We find that the EDSR$^\mathrm{off}$ pulse reduces the electron $T^\mathrm{H}_{\rm 2e}$ by a factor of two and matches $T^\mathrm{H}_{\rm 2ff}$ of the flip-flop qubit. Note that the exponent of the decay changes from $1.7(1)$ to $1.17(7)$, when applying EDSR$^\mathrm{off}$ pulses, which indicates a  change in the spectrum of the noise experienced by the electron spin \cite{yang2016quantum}.
	
	The previous experiments have demonstrated an effect of the EDSR$^\mathrm{off}$ pulse on the coherence of the electron qubit. In the following sections, we take a closer look on the underlying effect. We find that the EDSR$^\mathrm{off}$ drive tone causes a shift in resonance frequencies depending on its on duration and amplitude. Hence, for the measurements where we apply the magnetic and the electric tone simultaneously, the ESR drive becomes off-resonant which leads to the observed decay in the electron Rabi (\reffig{fig:ESR coherence}D). The $\pi$- and $\pi/2$- pulses used in the Ramsey and echo experiment deteriorate and decrease the spin-up proportion and refocusing properties for those measurements.
	
	\subsubsection*{ESR frequency shifts}
	\label{sec:esrstark}
	\begin{figure*}[htbp]
		\centering
		\includegraphics{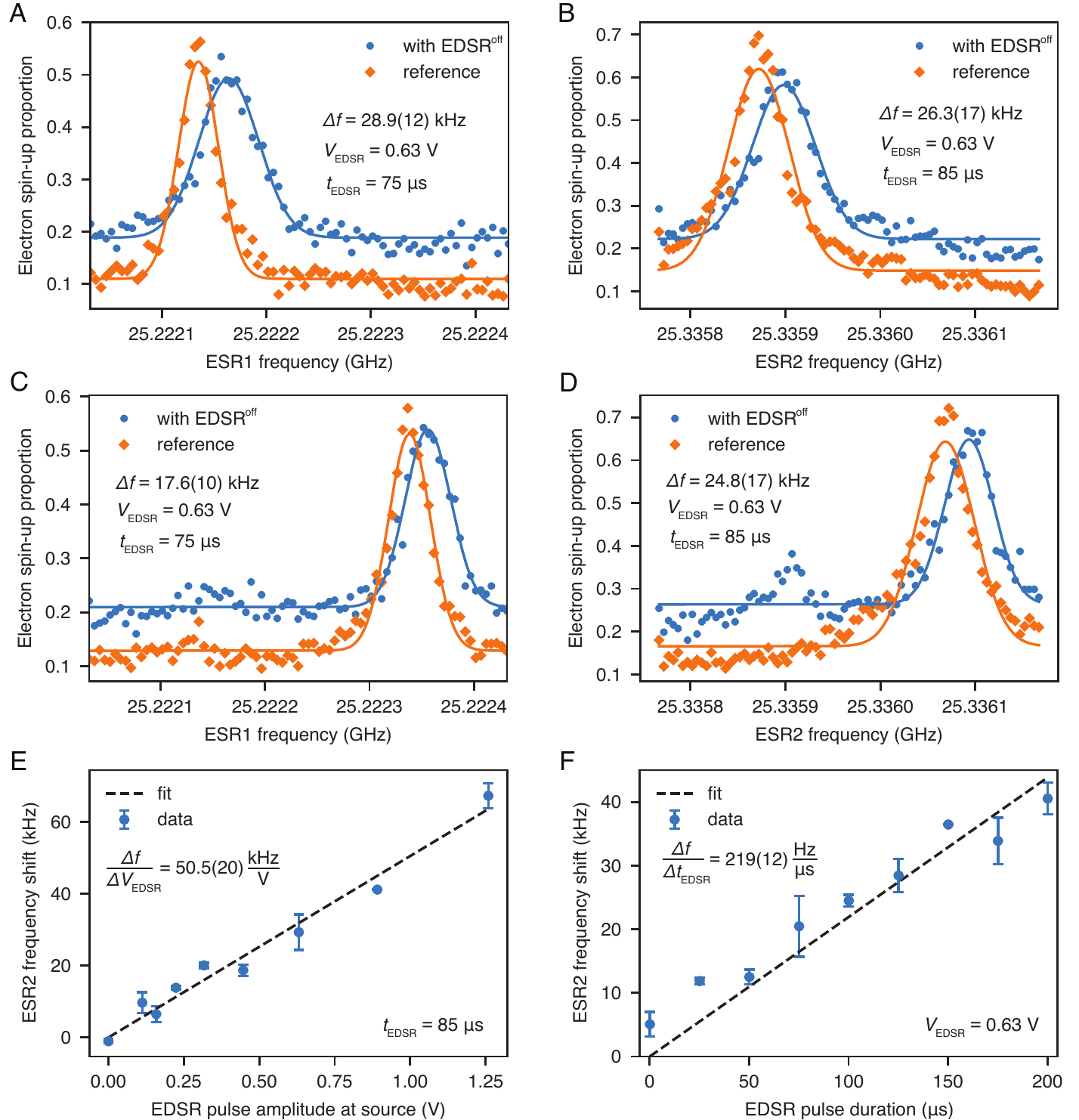}
		\caption{\textbf{ESR frequency shift.} (\textbf{A-D}) ESR spectrum scans averaged over multiple repetitions around the ESR1 (\textbf{A, C}) and ESR2 (\textbf{B, D}) resonances, measured with (blue dots) and without (orange diamonds) applying an EDSR$^\mathrm{off}$ pulse simultaneously with the ESR inversion pulse (see text). We show two resonance peaks (\textbf{A, C}) for ESR1 and (\textbf{B, D}) for ESR2 transitions corresponding to two detected configurations of nearby $^{29}$Si nuclear spins. The detected shifts in the resonance frequencies $\Delta f$ from applying the EDSR$^\mathrm{off}$ pulse are shown in the corresponding figures. (\textbf{E-F}) By changing the amplitude (\textbf{E}) and duration (\textbf{F}) of the EDSR$^\mathrm{off}$ pulse, we find a linear increase of the ESR2 frequency shift with $\Delta f / \Delta V_\mathrm{EDSR} = 50.5(20)$~kHzV$^{-1}$ (\textbf{E}) and $\Delta f / \Delta t_\mathrm{EDSR} = 219(12)$~Hz$\mu$s$^{-1}$ (\textbf{F}). For the measurement in (\textbf{F}), the EDSR$^\mathrm{off}$ pulse has 0.63~V amplitude at the MW source and its end is aligned to the end of the ESR pulse.}
		\label{fig:ESR stark shift}
	\end{figure*}
	
	To investigate the effect of an EDSR$^\mathrm{off}$ pulse on the ESR resonance frequency, we perform two interleaved ESR spectrum scans around the ESR1 and ESR2 resonances. The first scan is a regular ESR spectrum scan for reference; the second is taken while adding a 5~MHz off-resonant EDSR$^\mathrm{off}$ pulse at the same time as the ESR inversion pulse. The ESR inversion itself is performed using a $9\pi$ pulse instead of a simple $\pi$ pulse, to allow for a longer EDSR$^\mathrm{off}$ pulse and amplify its effects on the ESR resonances. The duration of these inversion pulses is 75~$\mu$s for ESR1 and 85~$\mu$s for ESR2 (different transmission for the two frequency ranges). For these measurements the magnetic field is set to $B_0 \approx 0.9$~T.
	
	Since most of the spectra are affected by $^{29}$Si flips (see~S4), we perform multiple repetitions and fit individual resonances with a Gaussian function before averaging (\reffig{fig:ESR stark shift}A-D). Compared to the reference peaks, we find that the resonance is at a higher frequency for both ESR1 and ESR2 frequencies when an EDSR$^\mathrm{off}$ tone is applied at the same time. 
	The ESR resonance is given by the Zeeman splitting and the hyperfine interaction $A$ as $f_\mathrm{ESR1/2} \approx \gamma_\mathrm{e} B_0 \mp A/2$. As both transitions are shifted towards higher frequencies, we conclude that the frequency shift is caused predominantly by a change in the electron $g_\mathrm{e}$-factor~\cite{laucht_electrically_2015}. The ESR frequency shifts appear to depend linearly on both the amplitude and the duration of the the off-resonance EDSR$^\mathrm{off}$ tone (\reffig{fig:ESR stark shift}E-F)
	
	\subsubsection*{NMR frequency shifts}
	\begin{figure*}[htbp]
		\centering
		\includegraphics{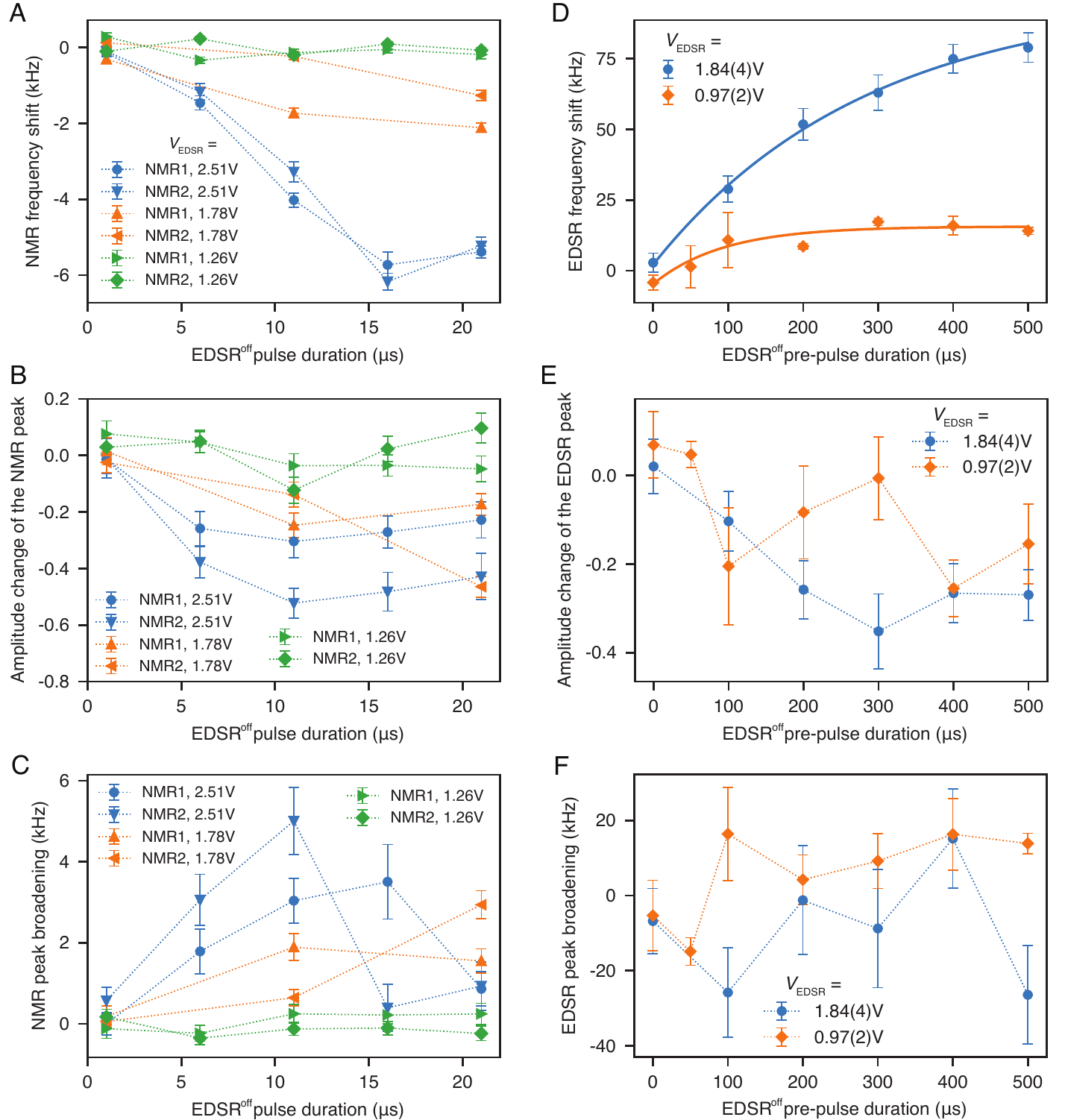}
		\caption{\textbf{NMR and EDSR frequency shifts.} (\textbf{A-C}) Frequency shift (\textbf{A}), amplitude change (\textbf{B}) and FWHM broadening (\textbf{C}) of the NMR1 and NMR2 resonance peaks in the spectrum scans for different durations of the EDSR$^\mathrm{off}$ pulse. We measure these dependencies for three amplitudes $V_\mathrm{EDSR} = 2.51$~V (blue), 1.78~V (orange) and 1.26~V (green) of the EDSR$^\mathrm{off}$ pulse at the MW source. \textbf{(D-F)} Frequency shift (\textbf{D}), amplitude change (\textbf{E}) and FWHM broadening (\textbf{F}) of the EDSR resonance peak in the spectrum scans for different durations of the EDSR$^\mathrm{off}$ pre-pulse (see text). We measure these dependencies for two amplitudes $V_\mathrm{EDSR} = 1.84(4)$~V (blue) and 0.97(2)~V (orange) of the EDSR pulses at the MW source. The EDSR frequency shifts are fitted with an exponential function $\delta f_\mathrm{EDSR} = \delta f_\mathrm{A} (1 - \mathrm{exp}(-t / \tau_{\rm sat})) + \delta f_0$, where the fit parameters are given in \reftab{tab:edsr}.}
		\label{fig:nmredsrstark}
	\end{figure*}
	
	Next we investigate the effects of the EDSR tone on the NMR spectra by measuring the NMR1 and NMR2 spectra and applying the off-resonant EDSR$^\mathrm{off}$ tone right before the NMR pulse. The extracted resonance frequency shifts, amplitudes and line widths as a function of the EDSR$^\mathrm{off}$ pulse duration are shown in \reffig{fig:nmredsrstark}. In \reffig{fig:nmredsrstark}A, we see that the resonances for NMR1 and NMR2 both shift to smaller values. This indicates a reduction in the hyperfine interaction as the NMR resonance frequency is given by $f_\mathrm{NMR1/2} \approx A/2 \pm \gamma_\mathrm{n} B_0$ (in the electric field range available to our system, the nuclear $\gamma_\mathrm{n}$ can be assumed constant). Compared to the ESR spectra, the NMR frequency shifts are at least five times smaller in absolute terms, confirming that the hyperfine shift is weaker than the $g_e$-factor shift affecting the ESR frequency. For low EDSR drive amplitudes, the NMR frequency shifts are not resolvable in the spectrum scans. As shown in \reffig{fig:nmredsrstark}B-C, the EDSR$^\mathrm{off}$ tone also changes the amplitude and the width of the NMR resonance.
	
	\subsubsection*{EDSR frequency shifts}
	\label{sec:EDSRshift}
	
	Changes of  the electron $g_\mathrm{e}$-factor caused by a strong electric drive tone also affect the resonance frequency of the flip-flop qubit. To quantify the EDSR frequency shifts, we measure the EDSR resonance frequency after applying an off-resonant EDSR$^\mathrm{off}$ pre-pulse and compare the spectrum to a reference measurement omitting the additional EDSR$^\mathrm{off}$ tone.
	\reffig{fig:nmredsrstark}D shows the EDSR frequency as a function of the EDSR$^\mathrm{off}$ pre-pulse duration for two different EDSR amplitudes. Contrary to the effect on the ESR and NMR resonances, we find that the frequency shift saturates at longer duration. The saturation time $\tau_{\rm sat}$ appears to change for different amplitudes of the EDSR$^\mathrm{off}$ tone. We fit the dependence with an exponential function of the form $\delta f_\mathrm{EDSR} = \delta f_\mathrm{A} (1 - \mathrm{exp}(-t / \tau_{\rm sat})) + \delta f_0$. The fit parameters are given in \reftab{tab:edsr}.
	
	Similar to the NMR frequency shifts, the amplitude of EDSR resonance peak also decreases with the pre-pulse duration (\reffig{fig:nmredsrstark}E). However, we do not find any clear duration dependence for the FWHM of the EDSR peak like we see for the NMR peak (\reffig{fig:nmredsrstark}C and F).
	
	\begin{table}[htbp]
		\centering
		\begin{tabular}{|c || c | c | c |} 
			\hline
			EDSR amplitude & $\delta f_\mathrm{A}$\,(kHz) & $\tau_{\rm sat}$\,($\mu$s) & $\delta f_0$\,(kHz) \\ [0.5ex] 
			\hline\hline
			1.84(4)~V & 94.8(50) & 284(34) & 2.1(18)  \\ 
			\hline
			0.97(2)~V & 20.4(32) & 93(37) & -4.7(28)  \\
			\hline
		\end{tabular}
		\caption{The table shows the parameters extracted from the fits to the EDSR frequency shift data. The fit model is given by $\delta f_\mathrm{EDSR} = \delta f_\mathrm{A} (1 - \mathrm{exp}(-t / \tau_{\rm sat})) + \delta f_0$.}
		\label{tab:edsr}
	\end{table}

\section{S6: Residual $^{29}$Si nuclear bath}
\label{sec:si29}

\begin{figure*}[htbp]
	\centering
	\includegraphics{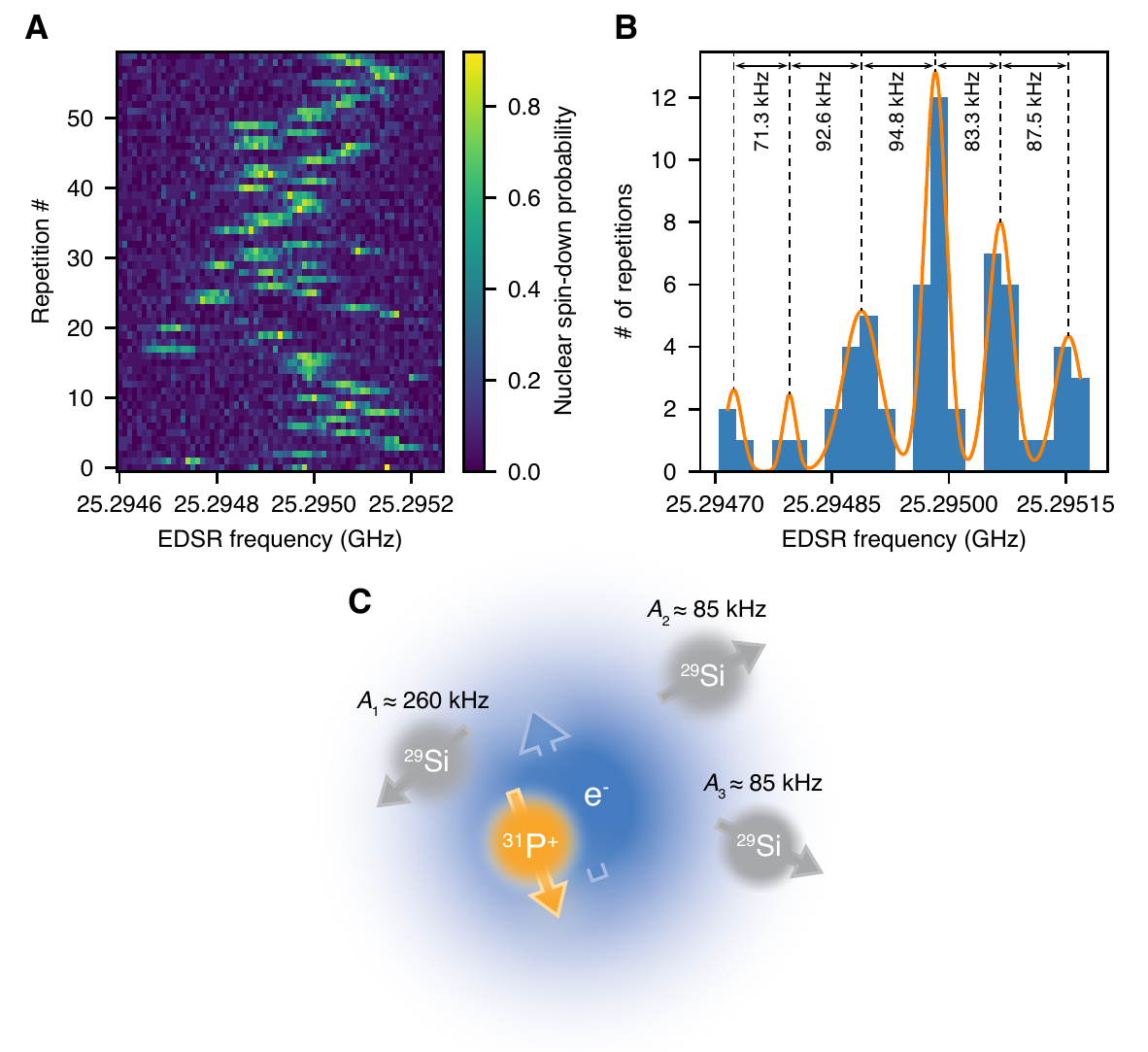}
	\caption{\textbf{$^{29}$Si nuclear spin flips.} \textbf{(A)} Tracking the EDSR resonance while applying a NMR pulse at the $^{29}$Si resonance frequency reveals discrete jumps in the flip-flop frequency. \textbf{(B)} The histogram shows the extracted EDSR resonance frequencies from \textbf{(A)}. We find six clusters of frequencies that are $\approx 19-55$~kHz wide (FWHM) and separated by $\approx 70-95$~kHz. \textbf{(C)} The spectra can be reproduced with at least three hyperfine coupled $^{29}$Si nuclei (see text for details).}
	\label{fig:si29}
\end{figure*}

The isotopically enriched $^{28}$Si epitaxial layer has a residual $^{29}$Si concentration of 730~ppm. With such value, one may expect of order 10 $^{29}$Si atoms within the Bohr radius of a $^{31}$P donor \cite{mkadzik2020controllable}, although only a small subset of them may possess a strong enough hyperfine coupling to the donor-bound electron to result in a visible effect on the resonance spectrum. We found that the donor under study in this paper is significantly coupled to (at least) three proximal $^{29}$Si atoms. The $^{29}$Si isotope has spin \nicefrac{1}{2} and a gyromagnetic ratio $\gamma_\mathrm{Si29}/2\pi = 8.465$~MHz/T and can couple to the donor-bound electron via the hyperfine interaction $A_\mathrm{Si29}$. In a semiclassical picture, this interaction can be considered as an additional magnetic field, which shifts any resonance frequency $f_\mathrm{res}$  depending on the coupling strength and the orientation of spins of the $^{29}$Si atoms:

\begin{equation}
	\label{eq:fsi29}
	f_\mathrm{res} = f_\mathrm{res,0} \pm \sum_i \frac{A_\mathrm{Si29,i}}{2}.
\end{equation}

Long-term EDSR spectrum measurements reveal that the EDSR resonance frequency randomly switches between 6 different values separated by $\approx 70-95$~kHz on a timescale that fluctuates between seconds and hours. To speed up the measurement and confirm that the frequency jumps are caused by the surrounding bath of $^{29}$Si, we apply a NMR $\pi$-pulse at the resonance frequency of the $^{29}$Si nuclei with the donor ionized, $\gamma_\mathrm{Si29} B_\mathrm{0}/2\pi \approx 7.644$~MHz ($B_0 = 0.9$~T for this experiment). In the absence of a hyperfine-coupled electron, this frequency is the same for all $^{29}$Si atoms and the NMR $\pi$-pulse will flip the $^{29}$Si spin configuration. In \reffig{fig:si29}\,A, we plot 60 EDSR spectra taken while applying a $^{29}$Si NMR pulse in between repetitions. We see that the EDSR resonance changes between most of the repetitions. A reference scan omitting the NMR pulses shows less frequency switching which indicates that the frequency jumps are caused by $^{29}$Si spin flips in the proximity of the qubit.

We extract the instantaneous EDSR resonance frequency by fitting individual spectra with a Gaussian function and show the histogram of the values in \reffig{fig:si29}\,B. We find a cluster of six EDSR frequencies separated by $\approx 70-95$~kHz and with a full-width-half-maximum of $\approx 19-55$~kHz. According to \refeq{eq:fsi29}, we conclude that the qubit is hyperfine coupled to at least three nearby $^{29}$Si atoms which should in principle result in $2^3=8$ different frequencies (see \reffig{fig:si29}\,C). We can reproduce the coupled system if we assume that two of the hyperfine couplings are within 10~kHz of each other, for instance $A_1 = 260(20)$~kHz and $A_2 \approx A_3 = 85(10)$~kHz.

The persistent frequency jumps require tracking of the instantaneous resonance frequency. Hence, we are forced to regularly perform frequency scans and update all relevant frequencies during and between measurements.

\section{S7: Gate Set Tomography and Randomized Benchmarking}

\begin{figure*}[htbp]
	\centering
	\includegraphics{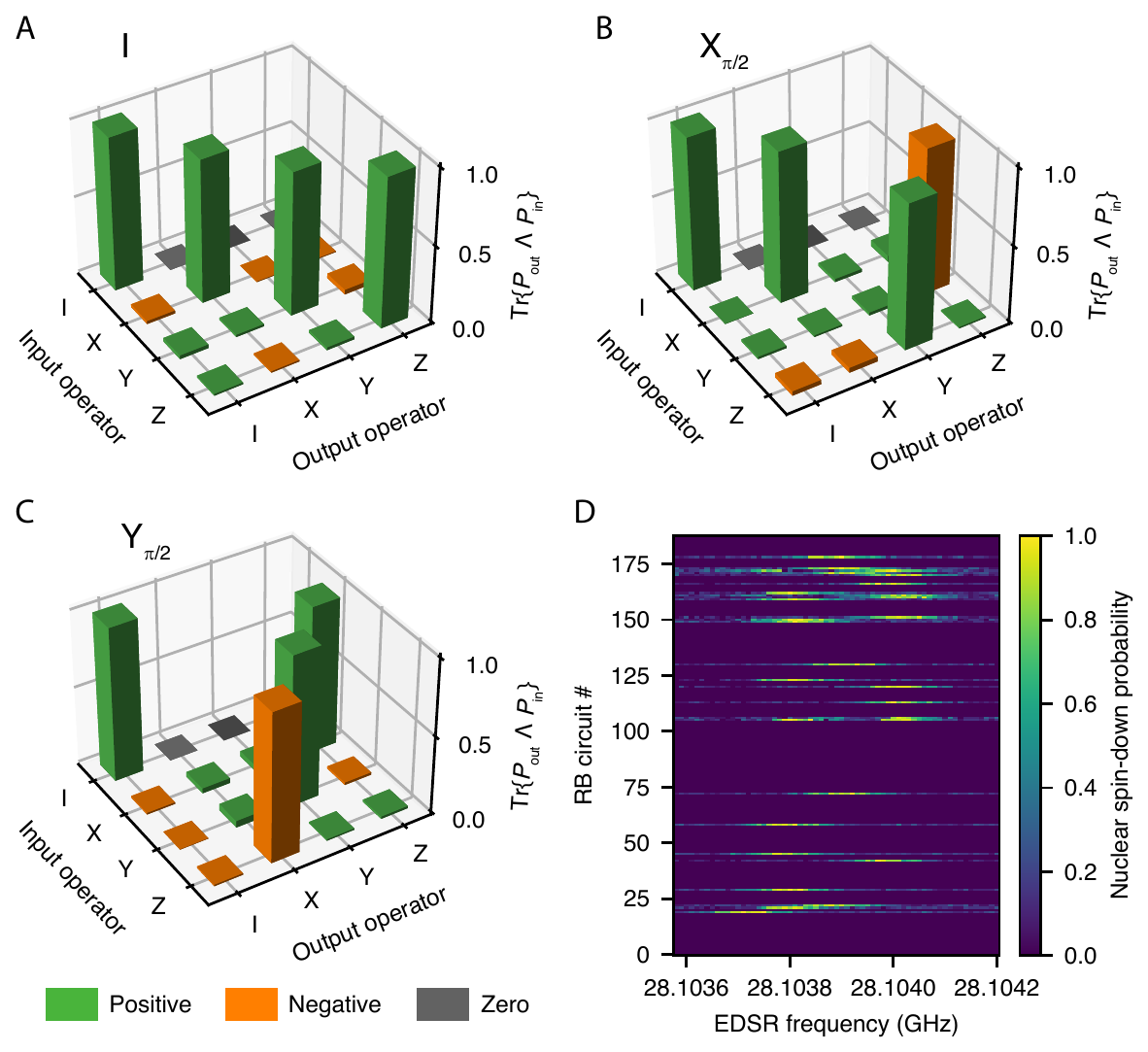}
	\caption{\textbf{GST process matrices and spectrum checks.} \textbf{(A), (B) and (C)} GST-estimated process matrices for the flip-flop qubit logic gates: \textbf{(A)} identity $I$, \textbf{(B)} $\pi/2$ rotation around $X$, $X_{\pi/2}$ and \textbf{(C)} and around $Y$, $Y_{\pi/2}$. \textbf{(D)} We track and update the EDSR resonance frequency before and after each measurement of the GST circuits to reduce the impact of $^{29}$Si spin flips. Blue lines indicate a stable configuration of the surrounding spins.}
	\label{fig:gstrb}
\end{figure*}

The gate set tomography (GST) protocol provides a detailed, calibration-free and self-consistent characterization of quantum gates~\cite{blume2013robust, greenbaum2015introduction, nielsen2021gate}. It quantitatively identifies gate errors and allows to correct some of them, e.g.~under- or over-rotation of the qubit. The gate set under investigation consists of a $X_{\pi/2}$, $Y_{\pi/2}$ and an identity $I$ gate. The $X_{\pi/2}$ and $Y_{\pi/2}$ gates are implemented as 90\degree-phase shifted, resonant EDSR $\pi/2$-pulses. The duration of 3.04~$\mu$s is limited by the maximum amplitude of the EDSR pulse of 2.59(6)~V that does not destabilize the device (see \refsec{ssec:dose}). The $I$ gate is implemented as a 1~$\mu$s long delay without applying any EDSR tone. The delay is chosen to be shorter than $T_{\rm 2ff}^* = 4.09(88)$~$\mu$s of the flip-flop qubit to limit the overall dephasing error during GST.

We measure the outcome of 448 GST circuits, each consisting of an initialization into \qdU state, a combination of gates $\in \{{X_{\pi/2},Y_{\pi/2},I}\}$ of varying length (up to maximum eight gates in the repeated germ sequences~\cite{nielsen2021gate, blume2017demonstration}) and a measurement of the nuclear spin \qD proportion. The initial circuit sequences characterize state preparation and measurement error of the gate set, whereas later circuits use error amplification techniques to map out the fidelity of the gates. The ideal measurement outcome of each circuit is a nuclear spin \qD proportion of either 0, 0.5 or 1; a deviation from these values indicates gate errors. 

Each GST circuit is repeated 100 times to collect output statistics, and the entire sequence of 448 circuits is repeated twice, thus yielding 200 measurement shots for each circuit. The measurement results are then fitted self-consistently by PyGSTi~\cite{pygsti}. The GST analysis determines single-qubit gate fidelities between $97.5\%-98.5\%$ (see Table~\ref{tab:S1}).

In addition, GST provides information about the contribution of different types of errors affecting the qubit control, which include coherent (under-/over-rotation of the qubit), stochastic (decoherence-like), affine (relaxation-like), and all other errors. While the majority of the infidelity is dominated by decoherence (between $36\%$ and $77\%$ of the total error), we still diagnose a small under-rotation of $0.8\degree - 1.31\degree$ for $X_{\pi/2}$ and $Y_{\pi/2}$ gates (see Table~\ref{tab:S1}). The affine errors are minor for our flip-flop qubit, as expected from the long flip-flop relaxation time shown in Sec.\,\ref{sec:T1}. For a further description of the error types that are analyzed by GST, see Refs.\,\cite{blume2022taxonomy, mkadzik2022precision}. GST also estimates the state preparation and measurement (SPAM) probability, i.e. the fidelity of the flip-flop qubit initialization into the \qdU state, yielding $F_\mathrm{SPAM} = 91.97\%$. This is in good agreement with the ENDOR initialization measurements in Subsec.\,\ref{sec:ninit}.

The reconstructed process matrices for the gate set are shown in \reffig{fig:gstrb}. The GST report also reveals model violations due to the presence of non-Markovian dynamics. Non-Markovianity is present in our system as a result of random $^{29}$Si spin flips and the shift in resonance frequency from applying high-power EDSR pulses (see \refsec{sec:si29} and \refsec{sec:stark}). To minimize the first effect, we check the EDSR resonance frequency before and after every GST sequence, and remeasure the sequence if the resonance frequencies don't match. However, we are still sensitive to $^{29}$Si spin flips within the measurement sequence itself and during the time it takes to upload the pulse sequence to the AWG. 

\begin{table*}[htbp]
	\centering
	\begin{tabular}{| c || c | c | c | c | c | c |} 
		\hline
		Gate & \begin{tabular}{@{}c@{}} Pulse duration\end{tabular} &
		\begin{tabular}{@{}c@{}} Rotation \\ angle\end{tabular} & \begin{tabular}{@{}c@{}} Coherent \\ errors\end{tabular} & \begin{tabular}{@{}c@{}} Stochastic \\ errors\end{tabular} & \begin{tabular}{@{}c@{}} Affine \\ errors\end{tabular} & \begin{tabular}{@{}c@{}} Fidelity \\ (average)\end{tabular} \\ [0.5ex] 
		\hline\hline
		$I$ & 1~$\mu$s delay & $0.0085\pi$ & 12\% & 77\% & 9.2\% & $97.5\%$  \\ 
		\hline
		$X_{\pi/2}$& 3.04~$\mu$s & $0.4955\pi$ & 33\% & 50\% & 11\% & $98.2\%$  \\
		\hline
		$Y_{\pi/2}$ & 3.04~$\mu$s & $0.4927\pi$ & 49\% & 36\% & 2.4\% &  $98.5\%$  \\
		\hline
	\end{tabular}
	\caption{This table shows partial results from the GST analysis for the flip-flop gate set. The errors are presented as a percentage of the total error.}
	\label{tab:S1}
\end{table*}

We additionally perform randomized benchmarking (RB), a comparatively simple, SPAM-insensitive characterization method that provides the average single qubit gate fidelity \cite{muhonen2015quantifying}. For a typical RB pulse sequence the qubit is first initialized in the \qdU state. Then we apply a random sequence of Clifford gates of length $m$ up to a maximum of 65. Before reading out the nuclear spin state, we apply a final Clifford gate that returns the qubit back to \qdU. Any deviation from \qdU implies errors in the Clifford gates. By varying the sequence length and measuring the decay constant, we can deduce the average gate errors of the Clifford gates. The Clifford gates are constructed from a combination of native $X_{\pi}, Y_{\pi}, X_{\pi/2}, Y_{\pi/2}$ gates. We used a set of gates from the open-source software PyGSTi where the average number of native gates per Clifford is $\approx 2.233$. 

In light of the results obtained earlier from GST, we adjusted the EDSR pulse duration of the $X_{\pi/2}, Y_{\pi/2}$ gates to account for the under-rotation detected in that experiment (compare \reftab{tab:S1} and \reftab{tab:S2}). As for the GST experiment, we sandwich every RB sequence between EDSR spectra and remeasure RB sequences if the resonance changes due to $^{29}$Si spin flips. We find an average Clifford gate fidelity $\mathcal{F}_\mathrm{C} = 96.4(5)\%$, which corresponds to an average native gate fidelity $\mathcal{F}_{\rm 1Q} = 98.4(2)\%$. The results obtained from RB are in good agreement with the GST fidelities.

In an attempt to account for PIRS effects, we adjusted the drive frequency with increasing RB pulse length to follow the measured exponential dependence (see \refsec{sec:EDSRshift}). Unfortunately, we did not find an increase in the average gate fidelity of those experiments in comparison to just using single-frequency sine pulses described above.  

\begin{table}[htbp]
	\centering
	\begin{tabular}{|c || c | c | c |} 
		\hline
		Gate & Pulse duration \\ [0.5ex] 
		\hline\hline
		$X$ & 6.13~$\mu$s  \\ 
		\hline
		$Y$ & 6.13~$\mu$s  \\ 
		\hline
		$\pm X_{\pi/2}$ & 3.073~$\mu$s  \\
		\hline
		$\pm Y_{\pi/2}$ & 3.087~$\mu$s  \\
		\hline
	\end{tabular}
	\caption{Native gates and their respective EDSR pulse durations used in the RB experiment.}
	\label{tab:S2}
\end{table}

	\section*{S8: High donor implantation dose}
	\label{ssec:dose}
	
	For a $^{31}$P$^{+}$ implantation fluence of $1.4\times10^{12}$~atoms/cm$^{2}$, we expect $\approx{40}$ donors in the implantation window at an average depth of $\approx$ 6.8(33)~nm below the SiO$_{2}$/Si interface. The high donor concentration increases the probability of finding a donor in a convenient location, but reduces the range of gate voltages that can be applied without affecting the charge state of nearby donors. 
	
	For a flip-flop qubit, the best gate performance and the most convenient multi-qubit coupling strategy are achieved in the high electric dipole regime, where an electron is significantly displaced from the donor atom \cite{tosi_silicon_2017}. Achieving this regime requires applying large voltage swings to the gates that control the donor potential.
	
\begin{figure}[htbp]
		\includegraphics[width=\columnwidth]{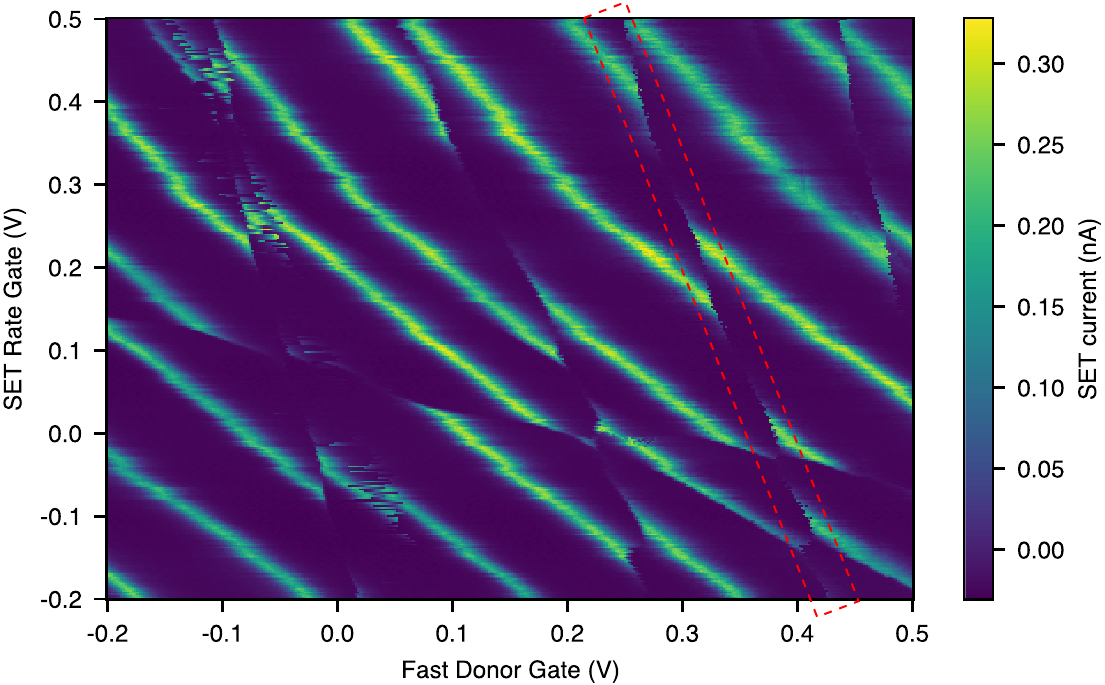}
		\caption{\textbf{Charge stability diagram.} The donor charge transition of the flip-flop qubit (red dashed rectangle is surrounded by additional charge transitions due to other nearby implanted donors.}
		\label{fig:charge_stab}
	\end{figure}

	In the present device, we prioritized having a high chance of finding a donor in a convenient location within the device. For this purpose we engineered 
	a $^{31}$P$^{+}$ implantation fluence of $1.4\times10^{12}$~atoms/cm$^{2}$, from which we expect $\approx{40}$ donors in the implantation window at an average depth of $\approx$ 6.8(33)~nm below the SiO$_{2}$/Si interface. The resulting charge stability diagram is shown in \reffig{fig:charge_stab}. It reveals numerous charge transitions -- breaks in the straight pattern of SET current peaks -- consistent with the large number of implanted donors. The charge transition corresponding to the donor used for the present experiment is shown encased in the red dashed rectangle. The presence of other donors limits the Fast Donor Gate voltage range around the transition to less than $\pm 200$~mV. Crossing other donor transitions would destabilize the electrostatic landscape of the device and perturb the operation of the qubit. Under such limitations, it was not possible to reach the high electric dipole regime.
	
	Future devices will adopt the deterministic single-ion implantation method recently demonstrated in our group, which allows for 99.85\% confidence in implanting a single donor \cite{jakob2022deterministic}. This will result is a clean stability diagram containing only one donor charge transition, and allow for the electrostatic tuning of the donor to the desired high-dipole regime \cite{tosi_silicon_2017}.
	
\end{document}